\documentstyle[12pt]{article}
\newcommand{\nn}{\nonumber}
\begin{document}
\begin{titlepage}
\begin{flushright}
Madras IMSc-91/33\\
IISc-CTS 7/91\\
\today\\
\end{flushright}
\vskip .10in
\begin{center}
 {\large\bf Three Dimensional Chern-Simons Theory as a Theory of Knots
and Links}
\vskip .10in
\end{center}
\centerline{\bf{R.K. Kaul$^{*}$}}
\centerline{\bf{Centre for Theoretical Studies, Indian Institute of Science}}
\centerline{\bf{Bangalore 560 012, India}}
\centerline{and}
\centerline{\bf{T.R.Govindarajan$^{\dag}$}}
\centerline{\bf{The Institute of Mathematical Sciences}}
\centerline{\bf{C.I.T. Campus, Taramani, Madras 600 113, India}}
\vskip .10in
\begin{center}
 {\large Abstract}\\
\end{center}
 Three dimensional SU(2) Chern-Simons theory has been studied as a
topological field theory to provide a field theoretic description of knots and
links in three dimensions.  A systematic method has been developed to obtain
the link-invariants within this field theoretic framework.  The monodromy
properties of
the  correlators of the associated Wess-Zumino SU(2)$_k$ conformal field
theory on a two-dimensional sphere prove to be useful tools.  The method is
simple enough to yield a whole variety of new knot invariants of which
the Jones polynomials are the simplest example.
\vspace*{8mm}
\begin{flushleft}
e-mail  \\
$*$: kaul@vigyan.ernet.in\\
       $\dag$: trg@imsc.ernet.in
\end{flushleft}
\end{titlepage}

\noindent {\bf1. Introduction}
\\
\par Topological quantum field theories have attracted a good deal of interest
in
recent years.  This started particularly with the field theoretic
interpretations of two important developements
in mathematics: Donaldson's theory for the integer invariants of smooth
4-manifolds in terms of the moduli spaces of SU(2)-instantons$^1$ and Jones
work
on knots in three dimensions$^2$. These were  developed by Witten some  years
ago$^{3,4}$.
Cohomological field theories involving monopoles of the three dimensions
have also been developed.$^5$
\\
 \par These applications of topological quantum field
theories reflect the enormous interest at present in building  a link between
quantum physics on one hand and geometry and topology of low dimensional
manifolds on the other.  It appears that the properties of low dimensional
manifolds can be very sucessfully unravelled  by relating them to infinite
dimensional manifolds of the fields.  This is done through the functional
integral
formulation of the such quantum field theories. In fact an axiomatic
formulation of topological quantum field theories has already been developed by
Atiyah$^6$.  This relates the functional integrals of
quantum field theory with of the notion of modular functors.
\\
\par Witten in his pioneering
work$^4$ has demonstrated that Jones polynomials of the knot theory are the
expectation values of the Wilson loop-operators in a three dimensional SU(2)
Chern-Simons theory where the fundamental representation of the gauge group
SU(2) lives on the knots.  The two  variable generalization$^7$ of the Jones
knot invariants are obtained as the expectation values of the Wilson loop
operators with N-dimensional representation living on the knots in an SU(N)
Chern-Simons theory.  The most important import of the Witten's formulation of
the knot theory is that it provides an intrinsically three dimensional
description of knots and links.  This field theoretic formulation is also
powerful
enough to study knots and links in any arbitrary three-manifold.
\\
\par The knot invariants have also been studied from the point of view of
exactly solvable models in statistical mechanics$^{8,9,10}$.  The intimate
connection that knot invariants have with the exactly solvable lattice
models has been exploited by Akutsu, Deguchi and Wadati to obtain a general
method
for constructing invariant polynomials for knots and links.  The Yang-Baxter
relation is an important tool in this development.  These authors in
particular have derived explicitly new knot invariants from a three-state
exactly solvable model.  These new invariants (both one variable and their
two variable generalizations) are indeed more powerful than the Jones
polynomials (and their two-variable generalizations).
\\
 \par The knot invariants
have also been studied from the quantum group point of view$^{11,12}$.
All these have an intimate connection with two dimensional conformal field
theory$^{13}$.
\\
 \par In this paper, following Witten, we shall study the
knot theory in terms of a topological quantum field theory.  The link
invariants are given by the expectation value of Wilson link-operators in
a Chern-Simons theory$^4$.  For definiteness we shall restrict ourselves to
the SU(2) gauge group.  Generalizations to other groups are straight forward
and will be discussed elsewhere.  By placing other than the doublet
representation of SU(2) on all the component knots of a link we obtain link
invariants other than those of Jones.  In particular, a triplet representation
on all the component knots leads to the one-variable invariants of Akutsu,
Deguchi and Wadati obtained from the three state exactly solvable model$^{10}$.
These invariants obey generalized Alexander-Conway skein relations containing
$n+2$ elements where $n$ is the number of the boxes in the Young tableaux
corresponding to the representation that lives on the knots.  For $n = 1$,
these relations are the standard Alexander-Conway skein relations that relate
Jones one-variable polynomials.  These were obtained within the field
theoretic framework by Witten in ref.[4].  This construction of Witten can
be extended to exhibit the generalized Alexander-Conway skein relations too.
After presenting a brief discussion of the non-Abelian Chern-Simons theory
in Section 2, we shall present a short derivation of the generalised
skein relations in Section 3.These
have also been obtained in ref.[14].
While these relations for $n = 1$ case (Jones polynomials) can
be recursively solved to obtain Jones polynomials for any knot or link,
those for $n \neq 1$  do not contain enough information to allow us to obtain
the new invariants for any arbitrary knot or link.  Therefore, there is
need to develop methods to obtain these invariants.  This is what we
attempt to do in this paper.  In Section 4, we shall present two new types of
recursion relations for links for arbitrary $n$.  In Section 5, invariants
for links obtained as the closures of braids with two strands, both parallely
oriented and anti-parallely oriented shall be obtained.  In Section 6, we
shall present some useful theorems for the functional integrals over
three-balls containing Wilson lines which meet the boundary at four points, two
incoming and two outgoing.  Next in Section 7, we shall develop the method
further and obtain building blocks for the calculation of the link
invariants.  These are used to calculate the invariants for the knots
upto seven crossing points as illustrations of the method in Section 8.  Then
we conclude with some remarks about generalizations of our results in Section
9.  Appendix A contains some useful formulae for the SU(2) quantum Racah
coefficients and the duality matrix which relates the four point  correllators
 of
the $SU(2)_k$ Wess-Zumino conformal field theory.  The field theoretic method
developed here can be used to rederive many nice properties of the Jones
polynomials $(n=1)$.  As an illustration of these, we shall present a proof
of the generalization of numerator-denominator theorem of Conway for
Jones polynomials in Appendix B.  In Appendix C we list some functional
integrals over
three manifolds containing Wilson lines ending on one or two boundaries, each
an $S^2$ with four
punctures,
obtained by the method developed earlier.  These
functional integrals are used to obtain the knot invariants presented
explicitly
in Section 8.
\newpage
\noindent {\bf2. Wilson Link-Operators in SU(2) Chern-Simons Theory in 3D}
\\
\par The metric independent action of Chern-Simons theory in a
three-manifold $M^3$ is given by
$$kS = {k\over 4\pi} \int\limits_{M^3} tr(AdA +{2\over{3}}  A^3) \eqno(2.1)$$
where $A$ is a matrix valued connection one-form of the gauge group G.  In
most of what follows we shall take the three manifold to be a
three-sphere, $S^3$.  The topological operators of this topological field
theory
are given in terms of the Wilson loop (knot) operators:
$$W_R [C] =  tr_R P exp \oint\limits_C  A \eqno(2.2)$$
for an oriented knot $C$.  These operators depend both on the isotopy type
of the knot as well as the representation $R$ living on the knot through
the matrix valued one-form $A_\mu=A_\mu^a T_R^a$  where $T^a_R$
are the representation matrices corresponding to the representation $R$ of
the gauge group.
\\
\par For a link $L =  (C_1, C_2, ..., C_s)$
made up of component knots $C_1, C_2, ... C_s$ carrying representations
$R_1, R_2, ..., R_s$  respectively, the Wilson link operator is defined
as
$$ W_{R_1 R_2 ...R_s} \ [L] = \prod_{i=1}^s
   W_{R_i}\ [C_i]  \eqno(2.3) $$
Unless indicated explicitly otherwise, in the following we shall place
the same representation $R$ on all the component knots of a link and hence
write the link operator  $W_{RR....R} [L]$  simply as  $W_R [L]$.  The
functional average of this link operator are the topological invariants
of this theory.These we define as follows :
$$ V_R [L]  =   \langle W_R [L]\rangle   \equiv
   {\int\limits_{M^3} [dA] \prod\limits^{s}_{i=1} W_R [C_i] e^{ikS}\over
     \int\limits_{S^3} [dA] e^{ikS}} \eqno.(2.4) $$
Clearly these invariants depend only on the isotopy type of the oriented
link $L$ and the representation $R$ placed on the component knots of the link.
This is so because both the integrand in the functional integral as well
as the measure$^{15}$ are independent of the metric of three-manifold on which
the
theory is defined.
\\
\par These functional integrals can be evaluated
by exploiting the connection of Chern-Simons theory on a three dimensional
manifold with boundary to corresponding Wess-Zumino conformal field
theory on the boundary$^4$.  Consider a three manifold $M^3$ with a set
of two-dimensional boundaries $ \Sigma^{(1)}, \Sigma^{(2)} .... \Sigma^{(r)}$.
Each of these boundaries $\Sigma^{(i)}$ may have a certain number of
Wilson lines carrying representations $R_j^{(i)} (j = 1,2 ...)$ ending or
beginning (at the punctures $P^{(i)}_j$) with representation $R^{(i)}_j (j =
1,2
..)$ on them.  With such a manifold we associate a state in the tensor
product of Hilbert spaces $\otimes^r$ $\cal H$$^{(i)}$ associated with
these boundaries, each with a certain number of punctures.  This state
then will represent the functional integral of the Chern-Simons theory
over such a manifold.  The dimensionality of each of these Hilbert
spaces $\cal H$$^{(i)}$ is given by the number of conformal blocks of the
corresponding Wess-Zumino conformal field theory on the respective
two dimensional boundaries with punctures $P_j^{(i)}$, j = 1,2 ...,
carrying the primary fields in representations $R_j^{(i)}$, j = 1,2 ...,
at these punctures.
\\
\par Exploiting this connection with the
Wess-Zumino conformal field theories, Witten has proved the following two
very simple theorems for the link invariants (2.4).
\\
\par {\bf Theorem 1:}  For the union of two distant (disjoint)
oriented links (i.e., with no mutual entanglements)$L_1, L_2$ carrying
representations
$R_1$ and $R_2$
$$ V_{R_1 R_2}  [L]  =  V_{R_1} [L_1] V_{R_2} [L_2] \eqno(2.5) $$
where    $ L = L_1\bigsqcup L_2$.
\\
\par {\bf Theorem 2:}  Given two oriented links $L_1$ and $L_2$ (Fig.1a)
their connected sum $L_1 \# L_2$, is obtained as shown in Fig.lb.  The
strands that are joined have to have the same representation living
on them and also the orientations on them should match.  Then the invariant
for the connected sum is related to the invariant for the individual
links as
$$ V_R [L_1 \# L_2]  =
   {V_R [L_1] V_R [L_2]\over V_R [\cup]}\eqno(2.6) $$
where $V_R [\cup]$ is the invariant of an unknot $\cup$ (i.e., $\bigcirc$).
\\
\par Besides these two theorems, within this field
theoretic framework Witten has also proved the Alexander-Conway skein relation
for the case
where the fundamental representation lives on all the components
of the link.  We shall now present a derivation of the generalized
Alexander-Conway skein relations for arbitrary representations living on the
links.
\vskip1cm
\noindent{\bf 3. Recursion Relations Among the Link Invariants}
\\
\par Before deriving recursion relations among the link invariants, let us
introduce a few useful definitions.
\\
\par Following, Lickorish and Millet$^{16}$,
we shall call a compact 3-dimensional submanifold in $S^3$ with
the boundary carrying a finite set of points marked by arrows as ``in"
or ``out", a room.  An inhabitant of the room is a properly embedded smooth,
compact oriented 1-manifold which meets the boundary of the room at the
given set of points with its orientation matching with the ``in" and ``out"
designations. Examples of rooms are shown in Fig.2.  Here 2(a) is a
three-ball with no markings on its boundary; 2(b) is a three-ball with two
marked points on its boundary, one ``in" and one ``out"; 2(c) is a
three-ball with four marked points on the boundary, two ``in" and two ``out".
Fig.2(d) shows a room with two inlets and two outlets with an example
of an inhabitant of the room drawn explicitly.\\  \par Now let us consider
a link $L_m$(A) as shown in Fig.3(a) made up of a room
\begin{picture}(10,7)
\put(5,1.5){\circle{7}}
\put(3,0){$A$}
\put(3,4){\vector(0,1){3}}
\put(7,4){\vector(0,1){3}}
\put(7,-4){\vector(0,1){3}}
\put(3,-4){\vector(0,1){3}}
\end{picture}
 with its
parallely oriented lower two strands containing a certain number of
half-twists m and then joined to the upper two strands of the room as shown.
The
half-twists are taken to be positive or negative if these are
right-handed or left-handed respectivily as shown in Fig.4.  Thus $m$ can be
$0,\pm 1,\pm 2,\pm 3 ...$ in the link $L_m(A)$ of Fig.3(a).  On each
component of this link, we place the spin $n/2$ representation $R_n$
of SU(2) given by Young tableaux with n horizontal boxes.
\\
\par Following Witten$^4$, we cut out a ball B$_1$ containing the room
\begin{picture}(10,7)
\put(5,1.5){\circle{7}}
\put(3,0){$A$}
\put(3,4){\vector(0,1){3}}
\put(7,4){\vector(0,1){3}}
\put(7,-4){\vector(0,1){3}}
\put(3,-4){\vector(0,1){3}}
\end{picture}
as well as all the twists in the lower two strands of this room as shown in
Fig.5(b).  The boundary S$^2$ of this ball is punctured at two ``in" and two
``out" points.  The normalized functional integral over this ball will
be
represented by $\psi_m(A)$.  The rest of S$^3$ containing the remaining
part of the link $L_m(A)$ is also a three-ball (B$_2$) with an
oppositely
oriented S$^2$ with four punctures (two ``in" and two ``out") as its
boundary as shown in Fig.5(b).  The normalized functional integral over
this ball is represented by $\overline \psi_0$.  These two normalized
functional integrals, $\psi_m(A)$ and  $\overline \psi_0$ are vectors in two
mutually dual Hilbert spaces, $\cal H$ and $\overline {\cal H}$,
associated  with the two oppositely oriented S$^2$'s forming the boundaries
of the two balls B$_1$ and B$_2$ respectively.  Each of this S$^2$ has
four punctures with representations R$_n$ attached to them.  The
dimensionality of each of these two vector spaces
is given by the number of conformal blocks of the correlator for four
primary fields, each in representation R$_n$, of the corresponding
Wess-Zumino SU(2)$_k$ conformal field theory on S$^2$.  The fusion rules of
this
conformal field theory are given by:  $R_n \otimes R_n=\oplus_{j=0}^{n}
R_{2j}$,
for  $k \geq 2n$.  (For  $k<2n$ the representations with $2j>k$ on the
right hand side are not integrable and hence are not to be included in the
fusion
rules).  We shall restrict our discussion to the case $k\geq 2n$.
However, the method developed here can also be used for $k<2n$.The
number of conformal blocks for the above four-point correlator is the
number of singlets in the decomposition of $R_n \otimes R_n \otimes $
$ R_n \otimes R_n $, which by the fusion rules is simply $n + 1$.
Hence the dimensionality of each of the vector spaces $ \cal H $ and
$ \overline {\cal H} $ is $(n+1)$ and normalized functional integrals
$ \mid \psi_m(A)\rangle$ and $\mid \overline \psi_0\rangle$ are $(n+1)$
dimensional vectors in these spaces, respectively.
\\
\par Now glueing back the
balls B$_1$ and B$_2$ gives us the original link $L_m(A)$.Thus the
invariant $V_n[L_m(A)]$ for this link can be represented through
the natural contraction of the vectors $\mid\psi_m(A)\rangle$ and
$\mid\overline\psi_0\rangle$ in the mutually dual Hilbert spaces,$\cal H$
and $\overline{\cal H}$, respectively:
$$V_n[L_m(A)] = \langle \overline\psi_0 \mid\psi_m(A)\rangle\eqno(3.1)$$
Here the subscript $n$ indicates that the representation $R_n$ with its
Young tableaux containing $n$ boxes is living on all the component knots.
\\
\par Now the vector $\mid\psi_m(A)\rangle$ represents normalized functional
integral over the ball $B_1$ with m half-twists in the inner two strands
as shown in the Fig.3(b).The operation of introducing a half-twist
in these two strands can be represented by an $(n+1)\times(n+1)$ matrix,
$\tilde B$
operating on these  vectors $\mid\psi_m(A)\rangle$ in the $(n+1)$ dimensional
Hilbert space $\cal H$.  Thus we may write
$$ (\tilde B)^j\mid\psi_{\it l}(A)\rangle = \mid\psi_{\it l+j}
    (A)\rangle, \ \ \ \ \     j= \pm1,\pm2 .... $$
The characteristic equation of the $(n+1)\times(n+1)$ half-twist monodromy
matrix $\tilde B$ can be written as
$$\sum \limits^{n+\ell+1}_{j=\ell}\tilde \alpha_{j-\ell}(\tilde B)^j = 0,
          \ \ \ \ \  \ell =0 ,    \pm1,\pm2, ... \eqno(3.2)$$
where the coefficients $\tilde\alpha_0, \tilde \alpha_1,.....\tilde
\alpha_{n+1} $ in terms of the eigen-values $\tilde \lambda_i, i =0,
1, .... n,$ of the matrix $\tilde B$ are
\begin{eqnarray}
\tilde\alpha_0 & = & (-1)^{n+1} \prod ^n_{i=0} \tilde \lambda_i,\nn  \\
\tilde \alpha_1 & =&  (-1)^n \sum ^n_{0\atop\scriptstyle{i_1\neq
i_2\neq....\neq i_n}}
\tilde\lambda_{i_1}\tilde\lambda_{i_2}....\tilde\lambda_{i_n}, \nn  \\
& \vdots &\nn \\  \tilde\alpha_{n-1}& =
&(-1)^2\sum^n_{0\atop\scriptstyle{i_1\neq i_2}}
\tilde\lambda_{i_1}\tilde\lambda_{i_2},\nn  \\
\tilde\alpha_n &=&\sum ^n_0\tilde\lambda_i ,\nn  \\
\hskip 2in\tilde\alpha_{n+1}\ & =&1 \hskip 3in (3.3)\nn
\end{eqnarray}
The eigen-values of the matrix $\tilde B$ are given by the monodromy
properties of the four-point correlator for the primary fields, all in
representation $R_n$, of $SU(2)_k$   Wess-Zumino conformal field theory
on $S^2$ (ref. 17):
$$\tilde\lambda_{j} = (-)^{n-j} exp [i \pi(2h_{n}-h_{2j})], \ \ \ \ \
j=0,1,\cdots
n\eqno(3.4)$$
where $ h_{j} ={{{j\over 2}({j\over2} +1)}/(k+2)}$ is the conformal weight of
the
primary field of spin $j/2$ representation $R_{j}$ of the $SU(2)_k$
Wess-Zumino conformal field theory.
\\
\par Applying  Eqn.(3.2) on the vector $\mid \psi_o(A)\rangle$ with no twists,
yields an equation amongst these vectors with successively increasing number of
half-twists as :
$$
\sum _{j=\ell}^{n+\ell+1}\tilde \alpha_{j-\ell} \mid \psi_{j}(A)\rangle =
0, \ \ \ \  \ell=0,\pm1,\pm2,\cdots \eqno(3.5)
$$
which in turn, using (3.1), yields a relation amongst link invariants
$L_{m}(A)$ as
$$
\sum_{j=\ell}^{\ell+n+1} {\tilde\alpha_{j-\ell}} V_{n}[L_j(A)] =
0, \ \ \ \  \ell=0,\pm1,\pm2,\cdots  \eqno(3.6)
$$
\par The link invariants may be given with respect to some reference framing.
Generally these are given in standard framing wherein each component knot of
the link has its self-linking number as zero.  Notice that the half-twist
matrix $\tilde B$ above does not preserve the framing.  Thus the various link
invariants in Eqn.(3.6) are not in the same framing.  In order to have all the
links in standard framing, the coefficients, $\tilde\alpha_m, m=0,1\cdots n+1$
in
(3.6) need to be multiplied by $exp(-2i{\pi}mh_n)$, respectively to cancel the
change in framing due to twisting.We implement this change in
Eqns.(3.2)-(3.6)  and for convenience  divide all of them  by an over all
factor  $exp(-2i\pi(n+1)h_n)$.This makes the effective
half-twist monodromy matrix for parallely oriented strands to be
$B=exp(2i{\pi}h_n)\tilde B$.
Its eigenvalues are given by :
$$
\lambda_{j} = (-)^{n-j} q^{(n(n+2)-j(j+1))\over 2} \ \ \ \ j=0,1,\cdots
n \eqno(3.7)
$$
with $q=exp[{2i\pi}/(k+2)]$, instead of those in Eqn.(3.4). Thus we rewrite the
final generalized Alexander-Conway skein recursion  relation
in the form of the following theorem:
\vskip1cm
\noindent{\bf Theorem 3:} The link invariants for links $L_{m}(A)$ of Fig.3 are
related as:
$$
\sum_{j=\ell}^{\ell+n+1} \alpha_{j-\ell}V_n [L_j(A)] = 0,
\ \ \ \ \  \ell = 0,\pm1,\pm2\cdots \eqno(3.8)
$$
where the coefficients $\alpha_{i}$ are given in terms of the eigenvalues of
the effective half-twist matrix $B$ above (Eqn.3.7) as :
\begin{eqnarray}
\alpha_0 & = & (-1)^{n+1} \prod ^n_{i=0} \lambda_i,\nn  \\
\alpha_1 & =&  (-1)^n \sum ^n_{0\atop\scriptstyle{i_1\neq
i_2\neq....\neq i_n}}
\lambda_{i_1}\lambda_{i_2}....\lambda_{i_n}, \nn  \\
& \vdots &\nn \\  \alpha_{n-1}& = &(-1)^2\sum^n_{0\atop\scriptstyle{i_1\neq
i_2}}
\lambda_{i_1}\lambda_{i_2},\nn  \\
\alpha_n &=&\sum ^n_0\lambda_i ,\nn  \\
\hskip 2in\alpha_{n+1}\ & =&1 \hskip 3in (3.9)\nn
\end{eqnarray}
This recursion relation has also been obtained in ref.14 within the field
theoretic formulation.
An earliar  derivation of this recursion relation within the frame-work of
exactly
solvable models is presented in ref.10.
\\
\par Now let us list some special cases of this relation :
{\bf(i)n=1:} Here the two eigen values of the effective monodromy
matrix $B$ are $\lambda_o =-q^{3/2}$ and $\lambda_1 = q^{1/2}$, so that
$\alpha_o = -q^{2}, \ \alpha_1 = q^{3/2} - q^{1/2}$ and $\ \alpha_2 = 1$.
These lead to the three-element recursion relation for the Jones polynomials as
$$
-qV_1[L_{\ell}(A)] + (q^{1/2} - q^{-1/2})V_1[L_{\ell+1}(A)] + q^{-1}
V_1[L_{\ell+2}(A)] = 0, \ \ \ \   \ell = 0,\pm1, \pm2 \cdots\eqno(3.10)
$$
{\bf(ii)n=2:} Here $\lambda_o=q^{4}, \ \lambda_1 = -q^3,\  \
\lambda_2=q$, and therefore, \ $\alpha_o=q^{8}, \  \alpha_1= -q^7 + q^5
-q^4, \ \alpha_2=-q^4 + q^3 - q$  and $\alpha_3=1$. This leads to the
four-element recursion relation:
\begin{eqnarray}
q^{4} V_{2}[L_{\ell}(A)] - (q^{3} - q+1)
V_{2}[L_{\ell+1}(A)]& &\nn \\
- (q^{-3} - q^{-1}+1) V_{2}[L_{\ell+2}(A)] &+&
q^{-4}V_2[L_{\ell+3}(A)] = 0, \ \ \ \  \ell=0, \pm1,
\pm2  \cdots (3.11)\nn
\end{eqnarray}
{\bf(iii)n=3:} \ Here $\lambda_{o}=-q^{15/2}, \ \
 \lambda_{1}=q^{13/2},
  \ \lambda_{2}=-q^{9/2}, \ \lambda_{3} = q^{3/2},$ and hence, $\alpha_{o} =
q^{20},
 \alpha_{1} = -q^{37/2} - q^{27/2} + q^{31/2} + q^{25/2}, \  \alpha_{2} =
-q^{14} +
q^{12} - q^{11} - q^{9} + q^{8} - q^{6}, \  \alpha_{3} = q^{15/2} - q^{13/2 } +
q^{9/2} - q^{3/2},
\alpha_{4} = 1.$  The recursion  relation here has five elements:
\begin{eqnarray}
q^{10} V_{3}[L_{\ell}] - (q^{17/2} - q^{11/2} + q^{7/2} - q^{5/2})
V_{3}[L_{\ell+1}] - (q^{4} - q^{2} + q + q^{-1} - q^{-2} + q^{-4})
V_{3}[L_{\ell+2}]& &\nn \\ -(q^{-17/2} - q^{-11/2} + q^{-7/2} - q^{-5/2})
V_{3}[L_{\ell+3}] + q^{-10} V_{3}[L_{\ell+4}] = 0,& &\nn \\  \ell = 0, \pm1,
\pm2\cdots \hskip .5in(3.12)& &\nn
\end{eqnarray}
\par Now we shall present two more recursion relations  which can be obtained
on the
same lines as above.
\vskip1cm
\noindent{\bf4. New Recursion Relations}
\\
\par Let us consider a link $\hat{L}_{2m}(\hat{A})$ as shown in Fig.5a, which
has even number of
half-twists 2m, in the oppositely oriented inner two strands.This link
can be obtained by glueing two balls $B_1$ and $B_{2}$ as shown
in Fig.5b, along their oppositely oriented boundaries.  The
normalized functional integrals over these two balls are represented by two
(n+1) dimensional vectors $\mid\chi_{2m}(\hat{A})\rangle$ and
$\mid\bar\chi_{o}\rangle$ in the two $(n+1)$
dimensional mutually dual Hilbert spaces associated with the two four-punctured
$S^{2}$'s forming the boundaries of these two balls respectively.  The
invariant for this link is given by the natural contraction of these two
vectors :
$$V_n[\hat{L}_{2m}(\hat{A})]= \langle \overline\chi_o
\mid\chi_{2m}(\hat{A})\rangle \eqno(4.1)$$
Let $\hat{B}$ be the $(n+1)\times(n+1)$ matrix introducing
half-twists in oppositely oriented inner two strands in the ball $B_1$.
Then
$$ (\hat{B}^2)^j \mid \chi_{2m} (\hat{A})\rangle  =  \mid
\chi_{2m+2j}(\hat{A})\rangle,
\ \ \ \ \ \ j = \pm1, \pm2,\cdots \eqno(4.2)$$
The eigen-values of this matrix  $\hat{B}$ introducing half-twists in
oppositely oriented strands on a four-punctured $S^2$, are given by
$$\hat{\lambda}_\ell = (-)^\ell q^{\ell (\ell+1)/2} \ \ \ \ \ \ \ \  \ell = 0,
1, \cdots  n \eqno(4.3)$$
The characteristic equation for $\hat{B}^2$ yields:
$$ \sum^{\ell +n+1}_{j=\ell} \hat{\alpha}_{j - \ell}
\hat{B}^{2j}= 0 \eqno(4.4)$$
and therefore
$$ \sum^{n + \ell + 1}_{j = \ell} \hat{\alpha}_{j-\ell}
\mid \chi_{2j}(\hat{A})\rangle  = 0 \eqno(4.5)$$
where $\hat{\alpha}_i,i=0,1, \cdots n+1 $
are related to the
eigen values of the  $\hat{B}^2$  matrix in the usual way given below in
Eqn.(4.7).Thus this yields us a recursion
 relation for the link invariants:
\vskip1cm
\noindent{\bf Theorem 4:} The invariants for the links in
Fig.5a are related as
$$\sum^{n + \ell + 1}_{j = \ell} \hat{\alpha}_{j -\ell}
V_n [{\hat L}_{2j}(\hat{A})] = 0, \ \ \ \ \ \ \ell =0, \pm1,\pm2 \cdots
\eqno(4.6) $$
where the coefficients  $\hat{\alpha}_i$  are given by
\begin{eqnarray}
\hat{\alpha}_0 & = & (-1)^{n+1} \prod ^n_{i=0} \hat{\lambda}^2_i,\nn  \\
\hat{\alpha}_1 & =&  (-1)^n \sum ^n_{0\atop\scriptstyle{i_1\neq
i_2\neq....\neq i_n}}
\hat{\lambda}_{i_1}^2\hat{\lambda}_{i_2}^2....
\hat{\lambda}_{i_n}^2, \nn  \\
& \vdots &\nn \\  \hat{\alpha}_{n-1}& =
&(-1)^2\sum^n_{0\atop\scriptstyle{i_1\neq i_2}}
\hat{\lambda}_{i_1}^2 \hat{\lambda}_{i_2}^2,\nn  \\
\hat{\alpha}_n &=&\sum ^n_0\hat{\lambda}_i^2 ,\nn  \\
\hskip 2in\hat{\alpha}_{n+1}\ & =&1 \hskip 3in (4.7)\nn
\end{eqnarray}
Here  $\hat{\lambda}_i$  are the eigenvalues of $\hat{B}$ matrix given in
Eq.(4.3).
\\
\par Now let us present a few special cases of this theorem:
{\bf(i)n=1:} Here the two eigen-values of the monodromy
matrix $\hat{B}$ are $\hat{\lambda}_0 = 1, \ \ \hat{\lambda}_1 = -q$.
Thus $ \hat{\alpha}_0  =  q^2 , \  \hat{\alpha}_1 = -(1+q^2),
 \ \hat{\alpha}_2 = 1$.  The recursion relations for  $\ell = 0,\pm 1,
\pm 2 \cdots ,$ then read :
$$ q V_1  [ \hat{\L}_{2\ell} (\hat{A}) ]  -  (q + q^{-1})
V_1 [ \hat{L}_{2\ell+2} (\hat{A}) ]  + q^{-1} V_1 [ \hat{L}_{2\ell+4}(\hat{A})]
=  0  \eqno(4.8) $$
{\bf(ii)n=2:} Here the eigenvalues of $\hat{B}$  are
$\hat{\lambda}_0 = 1, \ \hat{\lambda}_1 = -q, \
\ \hat{\lambda}_2 = q^3$
so that  $ \hat{\alpha}_0 = -q^8, \ \ \hat{\alpha}_1 = q^2 + q^6 + q^8,
 \ \hat{\alpha}_2 = -(1 + q^2 + q^6) $  and  $ \hat{\alpha}_3 = 1 $.  The
recursion relations  for $ \ell = 0,\pm1
\pm2....$,then read :
\begin{eqnarray}
- q^4 V_2[\hat{L}_{2\ell}(\hat{A})] + (q^{-2}+q^2+q^4)
V_2[\hat{L}_{2\ell+L}(\hat{A})]
&-& \nn  \\(q^{-4}+q^{-2}+q^2)V_2[\hat{L}_{2 \ell+4}(\hat{A})]
&+&q^{-4}V_2[\hat{L}_{2 \ell+6}(\hat{A})]= 0 \hskip2.2cm (4.9)\nn
\end{eqnarray}
\\
\par Similar recursion relations can be obtained for links shown in Fig.6 which
have odd number of half-twists in the oppositely oriented middle two
strands with the room
\begin{picture}(10,7)
\put(5,1.5){\circle{7}}
\put(3,0){$\hat A'$}
\put(3,4){\vector(0,1){3}}
\put(7,7){\vector(0,-1){3}}
\put(7,-4){\vector(0,1){3}}
\put(3,-1){\vector(0,-1){3}}
\end{picture}
as indicated in the Fig.6.  This
room is to be contrasted with the room
\begin{picture}(10,7)
\put(5,1.5){\circle{7}}
\put(3,0){$\hat A$}
\put(3,4){\vector(0,1){3}}
\put(7,7){\vector(0,-1){3}}
\put(3,-4){\vector(0,1){3}}
\put(7,-1){\vector(0,-1){3}}
\end{picture}
in Fig.5a.
An analysis as above then leads to the theorem:
\vskip1cm
\noindent{\bf Theorem 5:} For links $\hat{L}_{2m + 1} (\hat{A}')$
as shown in Fig.6, the link invariants are related as
$$
\sum_{j = \ell}^{n + \ell + 1} \hat{\alpha}_{j -\ell}V_n
[\hat{L}_{2j + 1}( \hat{A}')] =  0 \eqno(4.10)
$$
where $ \hat{\alpha}_i, \ \  i = 0, 1, \cdots  n + 1 $  are given by
equations (4.7) above.
\\
\par The recursion relation given in Theorems 3 - 5 do relate various link
invariants.  However, it is only for n = 1 (Jones Polynomials) that the
Alexander-Conway skein relation (3.10) along with the factorization property
of disjoint links given by Theorem 1 is complete.  That is, this can
recursively be solved for the link invariant of an arbitrary link.
For n = 2 and higher, this is not so.  Additional information is required
to obtain the link invariants.  Alternatively, methods need to be
developed to obtain the link invariants directly.  In the following sections
, we shall present one such method within the field theoretic
frame work.
\vskip1cm
\noindent{\bf 5. Invariants for Links Obtained as Closures of Two
Strand Braids}
\\
\par Consider the link $ L_m (A_1, A_2) $ obtained by glueing two balls along
their oppositely oriented boundaries $ S^2 $'s as shown in Fig.7.  The
ball $ B_1 $ contains the room
\begin{picture}(10,7)
\put(5,1.5){\circle{7}}
\put(3,0){$ A_1$}
\put(3,4){\vector(0,1){3}}
\put(7,4){\vector(0,1){3}}
\put(7,-4){\vector(0,1){3}}
\put(3,-4){\vector(0,1){3}}
\end{picture}
with the lower
two strands
oriented in the same direction and containing m half-twists.  The ball
$ B_2 $  contains the room
\begin{picture}(10,7)
\put(5,1.5){\circle{7}}
\put(3,0){$ A_2$}
\put(3,4){\vector(0,1){3}}
\put(7,4){\vector(0,1){3}}
\put(7,-4){\vector(0,1){3}}
\put(3,-4){\vector(0,1){3}}
\end{picture}
 .  The two boundaries are two
$ S^2 $ 's with four punctures each, two $``in''$  and  two  $ ``out''$.
The normalized functional integrals over these two balls are represented
by vectors  $ \mid \psi_m  (A_1) \rangle  $ and  $ \mid \overline\psi_0
(A_2)\rangle $ in two mutually dual (n+1) dimensional vector spaces
$ \cal{H}$  and  $ \overline{\cal{H}} $ respectively.  Let
$ \mid \phi_{\ell}\rangle , \ell = 0,1, \cdots  ,n $ be a complete
orthonormal set of eigen-vectors in  $\cal H$  of the braid matrix B which
introduces half-twists in the inner two parallel strands in  $B_1$ in
Fig. 7a :
$$ B \mid \phi_{\ell}\rangle = \lambda_\ell \mid \phi_\ell\rangle , $$
$$\lambda_\ell = (-)^{n-\ell} q^ {{n(n+2) -\ell(\ell+1)}\over2}, \ \ \ \
\ell = 0,1,2  \cdots ,n  \eqno(5.1)$$
The corresponding vectors in dual Hilbert space are denoted by
$ \mid \phi^\ell \rangle $  and the pairing of these is given by
$ \langle \phi^\ell \mid \phi_j \rangle = \delta^\ell_j $ .  The vectors  $\mid
\psi_0 (A_1)\rangle$  and  $ \mid \overline\psi_0 (A_2)
\rangle $  can then be expanded in terms of these bases :
\begin{eqnarray}
\mid \psi_0(A_1)\rangle & = &\sum^n_{\ell=0}\mu_ \ell (A_1) \mid\phi_\ell
\rangle \nn  \\
\hskip4.5cm \langle \overline\psi_0 (A_2)\mid& = & \sum^n_{\ell = 0} \mu_\ell
(A_2) \langle \phi^\ell \mid \hskip5cm (5.2) \nn
\end{eqnarray}
The vector with m half-twists $(m=0,\pm1,\pm2,\cdots )$ is
$$ \mid \psi_m (A_1)\rangle = \sum^n_{\ell = 0} \mu_\ell(A_1)
(\lambda_\ell)^m  \mid\phi_\ell \rangle \eqno(5.3)$$
The  invariant for the link $ L_m (A_1, A_2) $ of Fig.7, then can be
written as the contraction of the vector  $ \mid \psi_m (A_1) \rangle $  and
$ \mid \overline\psi_0 (A_2) \rangle $  as :
$$ V_n [L_m (A_1, A_2)] =  \sum^n_{\ell = 0} \mu_\ell (A_1)
 \mu_\ell (A_2)(\lambda_\ell)^m  \eqno(5.4) $$
Now if we can obtain a systematic method of finding the coefficients

$ \mu _\ell (A_1) $  and  $ \mu_\ell (A_2) $, we can write down these
link invariants. To do so let us take the two rooms to be
\begin{picture}(10,7)
\put(5,1.5){\circle{7}}
\put(3,0){$ A_1$}
\put(3,4){\vector(0,1){3}}
\put(7,4){\vector(0,1){3}}
\put(7,-4){\vector(0,1){3}}
\put(3,-4){\vector(0,1){3}}
\end{picture}
$=$
\begin{picture}(10,7)
\put(5,1.5){\circle{7}}
\put(3,0){$ A_2$}
\put(3,4){\vector(0,1){3}}
\put(7,4){\vector(0,1){3}}
\put(7,-4){\vector(0,1){3}}
\put(3,-4){\vector(0,1){3}}
\end{picture}
$=$
\begin{picture}(10,7)
\put(5,1.5){\circle{7}}
\put(7,-4){\vector(0,1){10}}
\put(3,-4){\vector(0,1){10}}
\end{picture}
 .Then the link  $ L_m (A_1,A_2) $ is
simply the link ${\cal L}_{m}$ obtained as the closure of m times twisted braid
of two parallely oriented strands as shown in Fig.8.Writing
simply  $\mu_\ell$  for  $ \mu_\ell$(
\begin{picture}(10,7)
\put(5,1.5){\circle{7}}
\put(7,-4){\vector(0,1){10}}
\put(3,-4){\vector(0,1){10}}
\end{picture}
) , the invariant
for this link is
$$ V_n [ {\cal L}_{m} ] =  \sum^n_{\ell = 0} \mu_\ell \mu_\ell
(\lambda_\ell)^m \eqno(5.5) $$
Now we need to determine $\mu_\ell$ .  Notice ${\cal
L}_0$ is simply two
unknots unlinked $ \cup \bigsqcup \cup $ ; $ {\cal L}_{\pm1} $ are
both one unknot $ \cup $ ; $ {\cal L}_{\pm 2} $ are right/left handed
Hopf links $(H, H^*)$ and $ {\cal L}_{\pm 3} $ are right/left handed
trefoil knots $(T, T^*)$.  Thus we may write
\begin{eqnarray}
V_n[{\cal L}_0 ] = &(V_n[\cup])^2
&= \sum^n_{\ell = 0} \mu_\ell \mu_\ell \nn
\\
  \hskip3cm V_n [{\cal L}_{\pm1}]= &V_n[\cup]
&= \sum^n_{\ell = 0} \mu_\ell \mu_\ell
(-)^{n-\ell} q^{\pm{{{n(n+2)}\over2}}
\mp {{\ell(\ell+1)}\over2}} \hskip 2.5cm (5.6) \nn
\end{eqnarray}
where we have used Theorem 1 to write  $V_n [\cup \sqcup
\cup]
=(V_n [\cup])^2$. From the second relation above, it is clear that  $V_n[\cup]$
is invariant under $q \rightarrow q^{-1}$ .This is so because unknot
has no chirality.These two equations can be solved
 recursively for various values of n. For n = 0, we have $ \mu_0 = 1 $.
Then using this, for n = 1, we obtain from these relations
$ \mu_1 = \sqrt{[3]} $.  For $n=2$, next we obtain $\mu_2 =\sqrt{[5]}$.
Thus in general  $ \mu_\ell = \sqrt{[2\ell+1]}$, where [m] is the
q-number defined as
$$
[m]={{q^{m/2} - q^{-m/2}}\over {q^{1/2} - q^{-1/2}}} \eqno(5.7)
$$
Using these in the first equation in (5.6) and the identity
$\sum \nolimits^n_{\ell =0}[2\ell+1]=[n+1]^2$ we have the knot
invariant for the unknot as
$$
V_n [\cup]  =  [n + 1] \eqno(5.8)
$$
This is not surprising, because the link invariant for two cabled knots
such as the two unlinked unknots obey the fusion rules.  That is, as
can be checked readily, the expression (5.8) for unknot obey the fusion
rule $V_{n_1}[\cup]V_{n_2}[\cup] = \sum^{min(n_1,n_2)}_{j =0}
V_{\mid n_{1}-n_{2}\mid + 2j}[\cup]$.
We can now put all this together in the form of a theorem :
\vskip1cm
\noindent{\bf Theorem 6:} For links ${\cal L}_{m}$ obtained as the closure of
a braid of two parallely oriented strands with m half-twists (Fig.8) the
link invariant is given by
$$
V_n[{\cal L}_{m}] = \sum ^n_{\ell = 0}(-)^{m(n-\ell)}
   q^{{m\over2}{(n(n+2)-\ell(\ell+1))}}[2\ell+1],
\ \ \ \ \   m = 0,\pm1,\pm2 \cdots  \eqno(5.9) $$
\\
\par A similar discussion can be carried through for the links of the type
shown
in Fig.9 and Fig.10.  The link  $ \hat{L}_{2m} (\hat{A_1}, \hat{A_2}) $ is
obtained by glueing the two connected balls $B_1$  and  $B_2$  as shown in
Fig.9.
Here we have two rooms
\begin{picture}(10,7)
\put(5,1.5){\circle{7}}
\put(3,0){$\hat A_1$}
\put(3,4){\vector(0,1){3}}
\put(7,7){\vector(0,-1){3}}
\put(3,-4){\vector(0,1){3}}
\put(7,-1){\vector(0,-1){3}}
\end{picture}
 and
\begin{picture}(10,7)
\put(5,1.5){\circle{7}}
\put(3,0){$\hat A_2$}
\put(3,4){\vector(0,1){3}}
\put(7,7){\vector(0,-1){3}}
\put(3,-4){\vector(0,1){3}}
\put(7,-1){\vector(0,-1){3}}
\end{picture}
 with 2m
half-twists in the oppositely oriented lower two strands of the first
room. The functional integrals over these two balls are again given by
vectors  $ \mid \chi_{2m} (\hat{A_1})\rangle $  and  $ \mid \overline \chi_0
(A_2)\rangle $ in mutually dual $n+1$ dimensional Hilbert spaces
associated with oppositely oriented four-punctured $S^2$. The invariant
of this link is given by natural contraction of these two vectors.  Here
the braid matrix $ \hat{B}$ introduces half-twists now in oppositely
oriented middle two strands of ball $ B_1$ , $\mid \chi_{2m}(\hat{A_1})
\rangle = \hat B^{2m}\mid \chi_0 (\hat{A_1})\rangle $.
Let $\mid \hat{\phi} _\ell \rangle $ be a
complete set of eigenfunctions, in vector space  $ \cal H $, of this
braid matrix  $ \hat{B} $,
$$\hat{B} \mid \hat{\phi}_{\ell}\rangle =  \hat{\lambda}_{\ell}
  \mid \hat{\phi}_{\ell}\rangle,\ \ \  \ \hat{\lambda}_\ell = (-)^\ell
  q^{{\ell(\ell +1)}\over2} , \ \  \ell = 0, 1, \cdots n  \eqno(5.10) $$
The corresponding basis in the
dual Hilbert space are denoted by $ \mid \hat{\phi^\ell}\rangle $, with their
natural contraction as $ \langle \hat{\phi}^\ell \mid \hat{\phi}_j\rangle =
{\delta}^\ell_j $.  Expand  $ \mid \chi_0 (\hat{A_1})\rangle $ and
$ \langle \overline \chi_0 (\hat{A_2}) \mid $ in these bases :
\begin{eqnarray}
\mid \chi_0 (\hat{A_1})\rangle &=& \sum^n_{\ell = 0}
\hat{\mu_\ell}(\hat{A_1})\mid \hat{\phi}_{\ell}\rangle
\nn \\
\hskip4cm \langle \overline \chi_0 (\hat{A_2})\mid & = & \sum^n_{\ell = 0}
\hat{\mu_\ell}
(\hat{A_2}) <\hat{\phi}^{\ell}\mid \hskip5cm (5.11)
\nn
\end{eqnarray}
Then the vector with 2m half-twists is
$$ \mid \chi_{2m} (\hat{A_1})\rangle = \sum \hat{\mu}_{\ell} (\hat{A_1})(\hat
{\lambda}_{\ell})^{2m} \mid \hat{\phi}_{\ell}
\rangle
\eqno(5.12) $$
so that the link invariant for the links of Fig.9 are given by
$$ V_{n}[\hat{L}_{2m}(\hat{A}_{1},\hat{A}_{2})] = \sum _{\ell=0}^n
\hat{\mu_{\ell}}
(\hat{A}_{1})\hat{\mu}_{\ell}(\hat{A}_2)(\hat{\lambda}_{\ell})^{2m}
\eqno(5.13) $$
In particular, if we take the two rooms to be
\begin{picture}(10,7)
\put(5,1.5){\circle{7}}
\put(3,0){$\hat A_1$}
\put(3,4){\vector(0,1){3}}
\put(7,7){\vector(0,-1){3}}
\put(3,-4){\vector(0,1){3}}
\put(7,-1){\vector(0,-1){3}}
\end{picture}
=
\begin{picture}(10,7)
\put(5,1.5){\circle{7}}
\put(3,0){$\hat A_2$}
\put(3,4){\vector(0,1){3}}
\put(7,7){\vector(0,-1){3}}
\put(3,-4){\vector(0,1){3}}
\put(7,-1){\vector(0,-1){3}}
\end{picture}
 =
\begin{picture}(10,7)
\put(5,1.5){\circle{7}}
\put(7,6){\vector(0,-1){10}}
\put(3,-4){\vector(0,1){10}}
\end{picture}
, then the link
$ \hat{L}_{2m}(\hat{A_1},\hat{A_2})$ simply becomes the link
$ \hat{\cal L}_{2m} $ obtained as the closure of two oppositely oriented
strands
with 2m half-twists as shown in Fig.10.  The invariants for these
links can be written as
$$ V_n [ \hat {\cal L}_{2m}] = \sum^n_{\ell = 0}\hat{\mu_\ell} \hat{\mu_\ell}
(\hat{\lambda_\ell})^{2m}  \eqno(5.14) $$
where  $ \hat{\mu_\ell} $ here refer to the room
\begin{picture}(10,7)
\put(5,1.5){\circle{7}}
\put(7,6){\vector(0,-1){10}}
\put(3,-4){\vector(0,1){10}}
\end{picture}
. Now, since
$\hat {\cal L}_0 = \cup \bigsqcup \cup $  and  $ \hat{\cal L}_{\pm 2} $
are right/left handed Hopf links $(H, H^*)$,  using $ V_n[\cup]=[n+1]$
and  $ V_n [H] = 1 + q + q^2 + \cdots q^{n(n+2)} $ and
$ V_n [H^*]= 1 + q^{-1} + q^{-2} + \cdots + q^{-n(n+2)} $  as obtained
using Theorem 6, we can solve for these  $ \hat{\mu_\ell} $'s successively
for $n = 0,1,2,\cdots $.This yields
$$ \hat{\mu}_{\ell}= {\mu}_{\ell} = \sqrt{[2\ell +1]}.$$Thus we collect these
results into a theorem.
\vskip1cm
\noindent {\bf Theorem 7:}  For links $ \hat{\cal L}_{2m} $ obtained as the
closure
of a braid made up of two oppositely oriented strands containing 2m half-
twists (Fig.10) the invariants are :
$$ V_n[\hat{\cal L}_{2m}] = \sum^n_{\ell =0}[2\ell + 1]
    q^{m\ell(\ell + 1)}, \ \ \ \ \ m = 0, \pm1, \pm2, \cdots  \eqno(5.15) $$
\\
\par After presenting these two simple theorems, we wish to develope this
method
further to obtain link invariants for more complicated links.
This we do in the next section.
\vskip1cm
\noindent{\bf 6. Some Useful Theorems for Link Invariants}
\\
\par Consider the two rooms $Q^{V}_{m}$ and $Q^{H}_{2p+1}$ with four markings
as
indicated in Fig 11a.  The first one has m half-twists vertically in the
parallely oriented strands and the latter has $2p+1$ half-twists horizontally
in
the oppositely oriented strands. The functional integral over the ball
containing these
rooms (Fig 11b) may be represented by vectors $\mid\psi(Q^{V}_{m})\rangle$ and
$\mid\psi(Q^{H}_{2p+1})\rangle$ respectively. The vector
$\mid\psi(Q^{V}_{m})\rangle$ is
obtained by applying the braid matrix $B$ (with eigenvalues
${\lambda}_{\ell}=(-)^{n-{\ell}}q^{{n(n+2)\over{2}}-{\ell(\ell+1)\over{2}}}$
and
eigenfunctions denoted by $\mid{\phi}_{\ell}\rangle , {\ell}=0,1,..n)$ of the
parallely
oriented central two strands m times on the vector
$\mid\psi(Q^{V}_{o})\rangle$.
This vector from the discussion of the previous section can be represented in
terms of the normalised eigenfunctions of $B$ as
$$
\mid{\psi}(Q_{0}^{V})\rangle=
\sum^{n}_{\ell=0}{\sqrt{[2\ell+1]}\mid{\phi}_{\ell}\rangle} \eqno(6.1)
$$
On the other hand, the vector $\mid{\psi}(Q^{H}_{2p+1})\rangle$ can be thought
of as
obtained by applying the braiding matrix $\hat{B}$ (with eigenvalues
$\hat{\lambda}_{\ell} = {(-)^{\ell} q^{{{\ell}(\ell+1)}\over2}}$ and
eigenfunctions
denoted by $\mid\hat{\phi}_{\ell}\rangle$, $\ell = 0,1,2,...n)$ of
anti-parallely
oriented side two strands in Fig11b on the vector
$$
\mid{\psi}(Q^H_0)\rangle = \sum^{n}_{\ell=0} {\sqrt{[2\ell+1]}
\mid\hat{\phi}_{\ell}\rangle} \eqno(6.2)
$$
which has been expanded in terms of the eigenfunctions
$\mid\hat{\phi}_{\ell}\rangle$
following the discussion of the previous section.The
basis referring to the first two strands on the left (or equivalently the last
two on
the right) in Fig.11b, $\mid\hat{\phi}_{\ell}\rangle$ and those refering to the
inner
two strands $\mid{\phi}_{\ell}\rangle$ are connected by the matrix $a_{j\ell}=
\langle{\phi}^{\ell}\mid\hat{\phi}_{j}\rangle$ where $\langle{\phi}^{\ell}\mid$
refers to the
basis with respect to the central two strands in the dual Hilbert space
obtained
by changing the orientation of the boundary $S^{2}$ of the ball.  This
discussion immediately allows us to write down the following theorem.
\vskip1cm
\noindent{\bf Theorem 8:} The functional integrals
$\mid{\psi}(Q^{V}_{m})\rangle$ and
$\mid{\psi}(Q^{H}_{2p+1})\rangle$ for the balls as shown in Fig 11b can be
written
in terms of the basis $\mid{\phi}_{\ell}\rangle$ refering to the parallely
oriented
middle two strands as :
$$ \mid{\psi}(Q^{V}_{m})\rangle = \sum ^{n}_{\ell=0}{\mu}_{\ell} (Q^{V}_{m})
\mid{\phi}_{\ell}\rangle
$$
$$
\mid{\psi}(Q^{H}_{2p+1})\rangle = \sum \mu_{\ell} (Q^{H}_{2p+1})
\mid{\phi}_{\ell}\rangle
$$
with
$$
\mu_{\ell}(Q^{V}_{m}) = (-1)^{m(n-\ell)} q^{{m\over2}(n(n+2)-\ell(\ell+1))}
\sqrt{[2\ell+1]}
$$
$$
\mu_{\ell}(Q^{H}_{2p+1}) = \sum ^{n}_{j=0}(-1)^{j} q^{(2p+1)
{j(j+1)\over2}} \sqrt{[2j+1]} \  a_{j{\ell}} \eqno(6.3)
$$
Here $\mu_{\ell}(Q^{H}_{2p+1})$ is obtained by first expanding in terms of the
basis $\vert{\hat\phi}_{\ell}\rangle$ referring to the first two strands and
then
changing the basis to $\vert\phi_{\ell}\rangle$ which refers to the middle two
strands, $\vert{\hat{\phi}}_{j}\rangle = \sum ^{n}_{\ell=0} a_{j\ell}
\vert\phi_{\ell}\rangle$.
\\
\par Similar discussion can be gone through with regard to the rooms ${\hat
Q}^{V}_{m}$,
${{\hat Q}^{H}_{2p}}$ and ${{\hat Q}^{H'}_{p}}$ as shown in Fig 12a and the
corresponding functional integrals for these balls (redrawn in Fig
12b),$\mid{\chi}(\hat{Q}_m^V)\rangle$,
$\mid \chi{({\hat Q}^{H}_{2p}}\rangle $ and
$\mid{\chi}{({\hat Q}^{H'} _p)}\rangle $ respectively. The middle two strands
here (Fig
12b) in all cases are oppositely oriented.  The basis
$\mid\hat{\phi}_{\ell}\rangle$ referring to these are eigen functions of braid
matrix $\hat{B}$ with eigenvalues $(-)^{\ell} q^{\ell(\ell+1)/2}$,
$\ell=0,1,...n$.  For the vector $\mid{\chi}({\hat Q}^{H}_{2p})\rangle$, the
braid
matrix with respect to the first two strands which are also oppositely oriented
has  the same eigenvalues. On
the other hand the braid matrix with respect to the first two strands in
$\mid{\chi}({\hat Q}^{H'}_{p})\rangle$ which are parallely oriented, has the
eigenvalues
$(-)^{n-\ell} q^{{{n(n+2)}\over2} -{{\ell(\ell+1)}\over2}}, \ell=0,1,...n$.
This
discussion, therefore, immediately leads to the theorem:
\vskip1cm
\noindent{\bf Theorem.9:}The functional integrals
$\mid{\chi}({\hat Q}^V_m)\rangle$,$ \mid{\chi}({\hat Q}^H_{2p})\rangle$ and
$\mid
\chi({\hat Q}^{H'}_{p})\rangle$ for the balls as shown in Fig 12b can be
written
in terms of the basis $\mid\hat{\phi}_{\ell}\rangle$ referring to the
anti-parallely oriented middle two strands as
$$
\mid\chi({\hat Q}^{V}_{m})\rangle  =  \sum^{n}_{\ell=0} \hat{\mu}_{\ell}
({\hat Q}^{V}_{m}) \mid \hat{\phi}_{\ell}\rangle
$$
$$
\mid\chi({\hat Q}^{H}_{2p})\rangle  =  \sum \hat{\mu}_{\ell}
({\hat Q}^{H}_{2p})\mid \hat{\phi}_{\ell}\rangle
$$
$$
\mid\chi({\hat Q}^{H'}_{p})\rangle  =  \sum \mu_{\ell}
({\hat Q}^{H'}_{p}) \mid \hat{\phi}_{\ell}\rangle
$$
where
$$
\hat{\mu_{\ell}}({\hat Q}^{V}_{m}) = (-)^{m\ell} q^{{m\ell(\ell+1)}\over2}
\sqrt{[2\ell+1]}
$$
$$
\hat{\mu_{\ell}}({\hat Q}^{V}_{2p}) = \sum^{n}_{j=0} q^{pj(j+1)} \sqrt{[2j+1]}\
  a_{j\ell}
$$
$$
\hat{\mu_{\ell}}({\hat Q}^{H'}_{p})  =  \sum^{n}_{j=0} (-)^{p(n-j)}
q^{p\left({{{n(n+2)}\over2} -{{j(j+1)}\over2}}\right)}\sqrt{[2j+1]} \
a_{j\ell} \eqno(6.4)
$$
Here again the matrix $a_{j\ell}$ relates basis with respect to the first two
strands (or equivalently last two strands) with the basis with respect to the
middle two strands in Fig 12b.
\\
\par The matrix $a_{j\ell}$ has very interesting properties.  To see this, let
us
glue two balls containing the rooms ${\hat Q}^H_{2m}$ and ${\hat Q}^H_{2p}$
respectively to obtain the link $\hat{L_{0}}(\hat Q^H_{2m}$, $\hat Q^H_{2p})$
as
shown in Fig 13a.  This link is the same as link ${\hat{\cal L}}_{2m+2p}$
obtained by the closure of the braid of two anti-parallel strands with $2m+2p$
half-twists discussed in the previous section.  Then the  invariant for
this link $V_{n}[\hat{L_{o}}{(\hat Q^H_{2m}, \hat Q^H_{2p})}]$ can be
written as
$$
\sum^{n}_{\ell=0} \hat{\mu_{\ell}}(\hat Q^H_{2m})
\hat{\mu_{\ell}}(\hat Q^H_{2p}) =  \sum^{n}_{\ell=0} q^{(m+p)\ell(\ell+1)}
\sqrt{[2\ell+1]}
$$
In this we substitute $\hat{\mu}_{\ell}(\hat Q^H_{2m})$ and
$\hat{\mu}_{\ell}(\hat Q^H_{2p})$ from Eqn 6.4.  Since this equation is
valid for arbitrary $m$ and $p$, we can equate the coefficients of various
powers of $q^{m}$, $q^{p}$.  This immediately yields :
$$
\sum^{n}_{\ell=0} a_{i\ell} a_{j\ell}  =  \delta_{ij}  \eqno(6.5)
$$
\par
On the other hand, if we compose two balls containing the rooms
${\hat Q}^H_{2m}$ and ${\hat Q}^V_{2p}$ as shown in Fig 13b, we have the
links $\hat{L_{o}}({\hat Q}^H_{2m}, {\hat Q}^V_{2p})$.  This link is the same
as the link $\hat{L_{o}}(\hat Q^H_{2p}, \hat Q^V_{2m})$ where $m$ and $p$
have been interchanged.  For the invariant for this link, we can write
$$
\sum^{n}_{\ell=0} \hat{\mu_{\ell}} (\hat Q^{H}_{2m}) \sqrt{[2\ell+1]} \
q^{p\ell(\ell+1)}  =  \sum^{n}_{\ell=0} \hat{\mu_{\ell}} (\hat Q^{H}_{2p})
\sqrt{[2\ell+1]}  \ q^{m\ell(\ell+1)}
$$
\noindent
where we have substituted for $\hat{\mu_{\ell}}(\hat Q^{V}_{2p})$ and
$\hat{\mu_{\ell}}(\hat Q^{V}_{2m})$ on the two sides of this equation from
(6.4).  Now if we substitute for $\hat{\mu_{\ell}}(\hat Q^{H}_{2m})$ and
$\hat{\mu_{\ell}}(\hat Q^{H}_{2p})$ also and since this relation is valid for
arbitrary $m$ and $p$, we have
$$
a_{i\ell}  =  a_{\ell i} \eqno(6.6)
$$
\noindent
Thus this matrix is symmetric and orthogonal.  Further in the link in Fig.13b
if we take
$p=0$, $\hat{L_{o}}(\hat Q^{H}_{m}$,$\hat Q^{V}_{o})$ is simply an unknot for
any value of $m$.  This implies
$$
\sum^{n}_{\ell=0} \hat{\mu_{\ell}} (\hat Q^{H}_{2m}) \sqrt{[2\ell+1]} \
  = [n+1]
$$
\noindent
Substitute for $\hat{\mu_{\ell}}(\hat Q^{H}_{2m})$ and equate coefficients of
various powers of $q^{m}$.  This leads to
$$
\sum^{n}_{\ell=0} \sqrt{[2\ell+1]} \  a_{\ell j} = [n+1] \delta_{jo} \eqno(6.7)
$$
\par
Next we consider the link $L_{o}(Q^{H}_{2p+1}, Q^{H}_{2m+1})$ obtained by
composing two balls containing the rooms $Q^{H}_{2p+1}$ and $Q^{H}_{2m+1}$
respectively as shown in Fig 14a.  This link is the same as the link
$\hat{\cal L}_{2m+2p+2}$ obtained as the closure of the braid with two
oppositely oriented strands with $2m+2p+2$ half-twists.  Hence its link
invaraint can be written as
$$
\sum^{n}_{\ell=0} {\mu_{\ell}}(Q^{H}_{2p+1})
{\mu_{\ell}}({Q^{H}_{2m+1}}) =  \sum^{n}_{\ell=0}
q^{(p+m+1)\ell(\ell+1)} [2\ell+1]
$$
\noindent
Here we substitute from Eqn(6.3) for $\mu(Q^{H}_{2p+1})$ and
$\mu(Q^{H}_{2m+1})$
and then equate coefficients of $q^{pi(i+1)}, q^{mj(j+1)}$ and
$q^{(p+m)\ell(\ell+1)}$ for various values of $i,j,\ell$. Thus we obtain again
the
orthogonality condition (6.5) for the matrix $a_{j\ell}$.  Similarly, the links
$L_{o}(Q^{H}_{\pm1}, Q^{V}_{m})$ obtained by glueing the two balls containing
the rooms $Q^{H}_{\pm1}$ and $Q^{V}_{m}$ respectively as shown in Fig 14b, are
the same as the links ${\cal L}_{m\pm1}$ obtained by the closure of  braids
with
two parallely oriented strands with $m\pm1$ half-twists for which link
invariants
are given by Theorem 6.  Thus
$$
\sum ^n_{\ell = 0} \mu_{\ell} (Q^H_{\pm 1})(-)^{m(n-\ell)}
q^{{m\over2}{(n(n+2)-\ell(\ell+1))}} [2\ell + 1]
= \sum ^n_{\ell = 0} [2\ell +1] (-)^{(m\pm1)(n-\ell)}
q^{{(m \pm 1)\over2}{(n(n+2)-\ell(\ell + 1))}}
$$
\noindent
Here substitute from Eqn(6.3) for $\mu_{\ell}(Q^{H}_{\pm1})$.  This yields
$$
\sum^{n}_{j=0} \sqrt{[2j+1]} (-)^{j} q^{{\pm}{j(j+1)\over{2}}} a_{j\ell} =
\sqrt{[2\ell+1]} (-)^{n-\ell} q^{{\pm}{1\over{2}}{(n(n+2)-\ell(\ell+1))}}
\eqno(6.8)
$$
Same equation would emerge if we had considered instead the link
$L_{o}{(Q^{V}_{2p+1},Q^{V}_{\pm1})}  \equiv \hat{\cal L}_{2p+2},\hat{\cal
L}_{2p}$ obtained by composing two balls containing the rooms $Q^{H}_{2p+1}$
and $Q^{V}_{\pm1}$ respectively.
\\
\par The matrix $a_{j\ell}$ satisfying conditions (6.5) - (6.8) is given in
terms of
quantum Racah coefficients for $SU(2)^{11-13}$
$$
a_{j\ell} = (-)^{\ell+j-n} \sqrt{[2j+1][2\ell+1]} \left( {\matrix { n/2 &
n/2 & j \cr n/2 & n/2 &\ell }} \right) \eqno(6.9)
$$
where the quantum Racah coefficients are as given in Appendix A.  We have also
presented some useful formulae for the Racah coefficients and the matrix
$a_{j\ell}$ in this Appendix.
\\
\par The fact that the matrix $a_{j\ell}$ relating the two bases is the quantum
Racah coefficient is not surprising.  The two bases referring to the side two
strands and the middle two strands attached to the four-punctured $S^{2}$'s
forming the boundaries of various balls considered above are related by duality
of the conformal blocks  for four-point correlations of
the corresponding $SU(2)_{k}$ Wess-Zumino conformal field theory on $S^{2}$.
The duality  matrix for these is indeed given by the quantum Racah
coefficients$^{13}$.
\\
\par It is clear that the Theorems 8 and 9 with the duality matrix $a_{j\ell}$
given
by Eqn.(6.9) can be used to construct a variety of link invariants.  We can
also
use these theorems for $n=1$ to rederive in the present framework many of the
results obtained for Jones
polynomial.  As an example we shall present a new
proof of the generalization of the numerator-denominator theorem of Conway for
the Jones polynomials derived by Lickorish and Millet in ref.16 within the
present framework in Appendix B.
\vskip1cm
\noindent{\bf 7. Building Blocks for Link Invariants}
\\
\par Although the results in Theorems 8 and 9 do allow us to write down
the link invariants for a class of links, these are not enough to study
the invariants for arbitrary links.  Now we shall attempt to develope
several other building blocks which will be useful in obtaining link
invariants for an arbitrary link.  For this, consider the three -
manifold $ S^3 $ from which several three-balls (say their number is
r) have been removed.  This yields a three-manifold with r boundaries,
each an $ S^2 $.  Wilson lines are placed inside this manifold such
that each boundary $ S^2 $ is punctured in four places, two ingoing
and outgoing.  The normalized functional integral over such a
manifold defines an operator in the tensor product of Hilbert spaces,
$ {\cal H}^{(1)} \otimes {\cal H}^{(2)} \otimes \cdots \otimes {\cal H}^{(r)}$
where $ {\cal H}^i $ is the (n + 1) dimensional Hilbert space associated
with the {\it{i}} th boundary.  This operator can be expanded in terms of a
convenient set of basis vectors corresponding to each of these Hilbert
spaces.
\\
\par For example, we have already argued that for a ball with
boundary as a four-punctured $ S^2 $, and Wilson lines as shown in Fig.15a,
the normalised functional integral is
$$ (Fig.15a) \ \   =  \sum^n_{\ell = 0} ~~ \sqrt{[2\ell + 1]} \mid
\hat\phi_{\ell}^{(1)}
\rangle      \eqno(7.1) $$

Here the functional integral  is multiplied by normalization
$ N^{-1/2} $ with N as the functional integral over boundaryless empty $ S^3 $.
The basis
$ \mid \hat\phi^{(1)}_{\ell}
\rangle $  of the (n+1) dimensional Hilbert space $ {\cal H} ^{(1)}$ associated
with the boundary in terms of which above expansion has been made refers to the
middle two (oppositely oriented) strands.  The half-twist matrix operating on
these strands is diagonal in
this basis with eigenvalues given by (5.10).  The signature $ \epsilon = + $
refers to the orientation of the boundary.  For opposite orientation,
$ \epsilon = - $, we expand in terms of the corresponding basis
$ \langle {\hat{\phi}}^{(1)\ell} \mid $  of the associated dual vector
space $ \overline {\cal H}^{(1)}$.
Thus the normalised functional integral over the
manifold in Fig.15b is
$$(Fig.15b) \ \  = \sum^n_{\ell = 0} \sqrt{[2\ell + 1]} \langle
 {\hat{\phi}}^{(1)\ell} \mid  \eqno(7.2) $$
\\
\par In contrast, the normalized functional integral over the ball with the
structure shown in Fig.16a
is:
$$ {\hat{\nu}}_1  = \sum^n_{j,\ell = 0}
\sqrt{[2\ell +1] [2j + 1]} \  a_{\ell j}\mid {\hat{\phi}}^{(1)}_j \rangle =
[n+1] \mid {\hat{\phi}}^{(1)}_0 \rangle\eqno(7.3) $$
\\
\par Next, let us consider a three-manifold with two boundaries, each a
four-punctured $ S^2 $, with four Wilson lines connecting them as indicated
in Fig.16b.  The normalized functional integral over this manifold can be
expanded in terms of the complete set of vectors $ \mid \hat\phi_{\ell}^{(1)}
\rangle $  and  $ \mid \hat \phi_{\ell}^{(2)}\rangle $ (referring to the middle
two strands in both cases) in the Hilbert spaces $ {\cal H}^{(1)}$ and
$ {\cal H}^{(2)}$ associated with the two boundaries :
$$ \hat \nu_2  =  \sum^n_{i,j=0} \hat{A}_{ij}
\mid {\hat{\phi}}^{(1)}_i \rangle \mid {\hat{\phi}}^{(2)}_j \rangle \eqno(7.4)
$$
\\
\par Similarly for the manifold with three boundaries
as shown in Fig.16c, each a four-punctured
$ S^2 $, we expand in terms of the bases  $ \mid \hat\phi^{(1)}_\ell \rangle $,
$ \mid \hat\phi^{(2)}_\ell \rangle,  \mid \hat\phi^{(3)}_\ell \rangle $ of
the three Hilbert spaces $ {\cal H}^{(1)}$,${\cal H}^{(2)}$,${\cal H}^{(3)}$
associated with these boundaries and referring to the inner two strands
in each case :
$$ \hat \nu_3 = \sum^n_{ij\ell} \hat{A}_{ij\ell}
\mid {\hat{\phi}}^{(1)}_i \rangle \mid {\hat{\phi}}^{(2)}_j \rangle
\mid {\hat{\phi}}^{(3)}_\ell \rangle \eqno(7.5)  $$
Here $ {\hat{\nu}}_3$ represents the functional integral normalized by
multiplying with
$N^{1/2}$ where  N is the functional integral  over the empty $S^{3}$.
\\
\par Now in Eqn.(7.4), the matrix-element $ \hat{A}_{ij} $ has to be
$ \delta_{ij} $.  This is obvious, because glueing the manifold shown in
Fig.16b onto the
manifolds in Figs.16a-c along an $ S^2 $ does not change any of these
manifolds.  That is, this functional integral (7.4) is an identity
operator.  On the other hand comparing the manifold in (7.5) with those
in (7.2) - (7.4) allows us to conclude that $ \hat{A}_{ij\ell}=
\sum_m a_{im}a_{jm}a_{\ell m}/ \sqrt{[2m + 1]} $ where
$a_{im}$ is the duality matrix.  Now glueing two manifolds of the type in
Fig.16c(Eqn.7.5)
with three boundaries along one each boundary with opposite orientations,
would yield a manifold with four boundaries.  Repeating this composition
several times then yields the following theorem :
\vskip1cm
\noindent{\bf Theorem 10:} The normalized functional integral for a manifold
with r boundaries, each an $ S^2 $, with Wilson lines as indicated
in Fig.17  expanded in terms of the bases  $ \mid \hat\phi^{(j)}_{\ell}
\rangle $ of the Hilbert spaces $ {\cal H}^{(j)}, j =1,2 \cdots r $
associated with the boundaries referring to the middle two strands in each
case,is
$$ \hat \nu_r =  \sum^n_{t = 0}
{{\prod^r_{j = 1} a_{{\ell_j}t}}\over{\left(\sqrt{[2t + 1]}\right)^{r-2}}}
\mid {\hat{\phi}}^{(j)}_{{\ell}_ j} \rangle \eqno(7.6) $$
Here the functional integral is normalised by  multiplying it by a factor $
(N^{1/2})^{r-2}$,
where N is the functional integral over boundaryless empty $ S^3 $.
\\
\par This functional integral may be acted upon by matrix  $ \hat{B}_{(j)},
j = 1,2 \cdots r  $ which introduces half-twists in the central two
strands of {\it {j}}th boundary :
$$
{\hat B_{(j)}} \vert \hat{\phi_\ell}^{(j)}\rangle = (-)^\ell
q^{\ell(\ell+1)\over2}
\vert {\hat \phi_\ell^{(j)}\rangle} \eqno(7.7)
$$
This matrix when operated $m_j$ times then will introduce $m_j$ half-twists in
the central two anti-parallel strands of the {\it {j}}th boundary leading to
factor
$(-)^{m_j\ell_j} q^{m_j\ell_j{(\ell_j+1)}/2}$ inside the summation on the
right hand side of Eqn.7.6.  On the other hand if we operate by the braiding
matrix $\hat B'_{(j)}$ which introduce half-twists in the first two (or
similarly in the last two) strands (again oppositely oriented) of the {\it
{j}}th
boundary, then
$$ ({\hat B'}_{(j)})^{m_j} \vert \hat \phi^{(j)}_\ell\rangle = \sum^n_{s=0}
(-)^{sm_j} q^{m_j s(s+1)/2} a_{\ell s} a_{sr}\vert
\hat \phi^{(j)}_r\rangle \eqno(7.8) $$
This way a whole variety of half-twists can be introduced in the diagram in
Fig.17 corresponding to Eqn.(7.6).  And then composing the resultant manifold
with functional integrals
of the type  in Eqns.(7.1-3) would lead to various link invaraints.
\\
\par In Eqn.(7.6), we have expanded the functional integral with respect to
the basis referring to the middle two strands on each of the boundary.
 The expansion with respect to the basis referring to the first two strands
(or equivalently the last two strands) is obtained by recognising that $
\sum^n_{\ell = 0}
a_{\ell t} \mid {\hat{\phi}}_{\ell}^{(j)} \rangle,
\ \  t = 0, 1, \cdots n $
form such a basis for each of the boundary.  Furthermore, the normalized
(by factor $(N^{1/2})^{r-2})$ functional integral with orientation on the
Wilson lines different from what are given in Fig.17 can also be obtained
in the same manner.  For example, the normalized functional intergal for the
manifold with r-boundaries, each
an $S^2$, with Wilson lines as indicated
 in Fig.18 is given by:
$$ \nu_r = \sum^n_{\ell = 0} {\mid \phi^{(1)}_{\ell}
\rangle \mid \phi^{(2)}_{\ell} \rangle \cdots \mid \phi^{(r)}_{\ell} \rangle
\over
\left(\sqrt{[2\ell + 1]}\right)^{r-2}}    \eqno(7.9)  $$
where we have now expanded this functional integral with respect
to the basis referring to the first two parallely oriented strands (or
equivalently the last two strands) on each of the boundaries.  Again, here
this functional integral
is multiplied by normalization $(N^{1/2})^{r-2}$.
Now the half-twists in the first two strands of the {\it {j}}th boundary which
have the same orientation are introduced through the matrix $B_{(j)}, j = 1, 2,
\cdots r $ with
$$ B_{(j)} \mid \phi^{(j)}_{\ell} \rangle = (-)^{n-\ell}q^{{1\over2} {(n(n+2) -
\ell(\ell+1))}} \mid \phi^{(j)}_{\ell} \rangle \eqno(7.10)$$
so that $ m_j $ half-twists would mean a factor $ (-)^{{m_j}(n-\ell)}
q^{{{m_j}\over2}(n(n+2) - \ell(\ell + 1))}$ inside the summation in the
right-hand side of Eqn.(7.9).  The $ m_j $ half-twists in the last two
strands (again parallely oriented) also introduce the same factor because
$ \mid \phi^{(j)}_{\ell} \rangle $ are also eigen-vectors with the same
eigenvalues of the braid matrix for these two strands.  On the other hand
for $ m_j $ half-twists in the middle two strands (which are oppositely
oriented) of the {\it {j}}th boundary are introduced through the braid matrix
$ {B'} _{(j)} $ :
$$(B'_{(j)})^{m_j}\mid \phi^{(j)}_{\ell} \rangle = \sum^n_{s = 0}
(-)^{m_j s} q^{{m_j}{s(s+1)\over2}} a_{\ell s} a_{sr}\mid \phi^{(j)}_r \rangle
\eqno(7.11)   $$
\\
\par Using the method described here, the functional integral over various
three-manifolds containing Wilson lines can be constructed.  As another
example, the normalized functional integral over the manifold as indicated
in Fig.19  containing 2r + 2 boundaries,  $ r = 0, 1, 2, \cdots $ can be
expanded as :
$$ (Fig.19) \ = \ \left({1\over{[n+1]}}\right)^r\sum^n_{i_{\ell =0}}\mid
\phi_{i_1}^{(1)}\rangle
\mid \phi_{i_1}^{(2)} \rangle \bullet \mid \phi_{i_2}^{(3)} \rangle
\mid \phi_{i_2}^{(4)} \rangle \bullet \ \ \cdots \mid \phi_{i_{r+1}}^{(2r+1)}
\rangle
\mid \phi_{i_{r+1}}^{(2r+2)} \rangle \eqno(7.12)$$
Here $ \mid \phi^{(j)}_{\ell} \rangle $ are the basis of the Hilbert space
${\cal H}^{(j)}$ associated with the {\it
{j}}th boundary and referring to the first
two strands or equivalently the last two strands (oriented in the same
direction) attached to that boundary.
\\
\par It should be noted that in
construction of all these functional integral that half-twists are to be
introduced in the strands attached to a four-punctured $ S^2 $ though
braid matrices which have eigenvalues $ (-)^{n-\ell}  q^{{1\over2}{(n(n+2) -
\ell(\ell+1))}},  \ell=0,1,\cdots n$ if the strands are oriented
in the same direction and  $ (-)^{\ell}  q^{{1\over2}{\ell(\ell+1)}},
\ell =0,1, \cdots n $  if the strands are oriented in opposite directions.
A number of useful building blocks can be constructed in this manner which then
readily yield the link invariants by appropriate composing of the
three-manifolds.  In Appendix C, we have listed some of these building blocks
for reference.  They have been used to calculate the knot invariants discussed
in the next section.
\\
\par The method described above can further be
generalized to obtain building blocks representing normalized functional
integrals over three manifolds with $ S^2 $ boundaries which are punctured
by Wilson lines at 6,8,10 ... points.   The dimensionality of the Hilbert
spaces associated with such boundaries is again given by the number of
conformal blocks of 6-point, 8-point, 10-point, .... correlators of the
associated $ SU(2)_k $ Wess - Zumino model on these boundaries.  The
duality matrix relating different bases of each of these Hilbert spaces
associated with $S^2$ boundaries
punctured at $2m+2$ points with $m=0,1,\cdots $
are given in terms of  quantum 6mj symbols.
\vskip1cm
\noindent{\bf 8. Explicit Calculation of Knot Invariants}
\\
\par Here we shall present the knot invariants for some knots
calculated from the building blocks listed in Appendix C as illustrartions.  It
is possible
to obtain the  invariants for all the knots and
links listed in, for example,
Rolfsen's book or the book by Burde and Zieschang$^{18}$.  However below we
shall
list the answers only for knots upto seven crossing number.  For a knot these
invariants are unchanged if its orientation is reversed.For  the
mirror reflected knot the invariants are given by the conjugate expressions
obtained
by replacing $q$ by $q^{-1}$.
\\
\par
The invariants $V_n[L]$ for knots with seven crossing number listed in Fig.20
are given by
\begin{eqnarray*}
0_1: V_n & = & [n+1]\\
3_1: V_n & = & \sum^n_{\ell=0} [2\ell+1](-)^{(n-\ell)}
q^{-{3\over2}{(n(n+2)-\ell(\ell+1))}}\\
4_1: V_n & = & \sum^n_{\ell,j=0} \sqrt{[2j+1][2\ell+1]} a_{j\ell}
q^{\ell(\ell+1)-j(j+1)}\\
5_1: V_n & = & \sum^n_{\ell=0} [2\ell+1] (-)^{n-\ell} q^{{-5\over2}
(n(n+2)-\ell(\ell+1))}\\
5_2: V_n & = & \sum^n_{j,\ell=0} \sqrt{[2j+1][2\ell+1]} a_{j\ell}(-)^j
q^{n(n+2)-\ell(\ell+1)+{3\over2}j(j+1))}\\
6_1: V_n & = & \sum^n_{j,\ell=0} \sqrt{[2j+1][2\ell+1]} a_{{\ell}{j}} q^{\ell
(\ell +1)-2j(j+1)}\\
6_2: V_n & = & \sum^n_{i,j,\ell=0} \sqrt{[2i+1][2j+1]} a_{{\ell}{i}}
a_{{\ell}{j}}
(-)^{n-\ell-j} q^{-{3\over2}(n(n+2)-j(j+1))-{{\ell(\ell+1)}\over2}+i(i+1)}\\
6_3: V_n & = & \sum^n_{j,\ell,r,s} \sqrt{[2j+1][2s+1]}a_{j\ell} a_{{\ell}{r}}
a_{rs}
(-)^{\ell+r} q^{-j(j+1)+s(s+1) + {{\ell(\ell+1)}\over2} - {{r(r+1)}\over2}}\\
7_1: V_n & = & \sum^n_{\ell=0} [2\ell+1] (-)^{n-\ell}
q^{-{7\over2}(n(n+2)-\ell(\ell+1))}\\
7_2: V_n & = & \sum^n_{j,\ell=0} \sqrt{[2j+1][2\ell+1]} a_{j\ell} (-)^j
q^{-n(n+2)+\ell(\ell+1)-{5\over2}j(j+1)}\\
7_3: V_n & = & \sum^n_{j,\ell=0} \sqrt{[2j+1][2\ell+1]} a_{j\ell} (-)^j
q^{{3\over2}j(j+1) + 2n(n+2) - 2\ell(\ell+1)}\\
7_4: V_n & = & \sum^n_{j,\ell,r}\sqrt{[2j+1][2r+1]} a_{j\ell} a_{r\ell}
(-)^{n-j-\ell-r} q^{{{n(n+2)}\over2} - {{\ell(\ell+1)}\over2} + {3\over2}
j(j+1) +
{3\over2} r(r+1)}\\
7_5: V_n & = & \sum^n_{ij\ell} \sqrt{[2i+1][2j+1]} a_{{\ell}{i}} a_{{\ell}{j}}
(-)^{n-j} q^{-{5\over2}n(n+2) + {3\over2}j(j+1)+ i(i+1)- \ell(\ell+1)}\\
7_6: V_n & = & \sum^n_{j\ell r s} \sqrt{[2j+1][2s+1]} a_{j\ell} a_{{\ell}{r}}
a_{rs}
(-)^{\ell} q^{-n(n+2)+j(j+1)-{{\ell(\ell+1)}\over2} + r(r+1) - s(s+1)} \\
7_7: V_n & = & \sum^n_{j \ell r s p} \sqrt{[2j+1][2p+1]}a_{j\ell} a_{{\ell}{s}}
a_{sr} a_{rp}
(-)^{n-r-s-\ell} q^{{{-n(n+2)}\over2} + j(j+1)-{{\ell(\ell+1)}\over2}-
{{r(r+1)}\over2} + {{s(s+1)}\over2}}
\end{eqnarray*}
Here $a_{j\ell}$ is the duality matrix given in Appendix A.
\\
\par Using the building blocks for normalized functional integrals over three
manifolds with 4  and higher punctures on their $S^2$ boundaries, the
above calculations can be extended to obtain the invaraints for the whole
tables
of knots and links given in ref.17 in a straight forward manner.
\\
\par Using the explicit representation for the
duality matrix $a_{jl}$ for $n=1$ as given in
Appendix A, the above expressions can easily be
seen to yield the Jones one-variable
polynomials$^2$ for these knots.On the other
hand,with the help of explicit representation
for $a_{jl}$for $n=2$ as given in Appendix A,we
obtain the polynomials$^{19}$ calculated explicitly by
Akutsu,Deguchi and Wadati from the three-state
exactly solvable model$^{10}$.
\vskip1cm
\noindent{\bf 9.Concluding Remarks}
\\
\par Following Witten$^4$, here we have studied the $SU(2)$ Chern-Simons theory
in three dimensions as a theory of knots and
links.  A systematic method has been developed to obtain link invaraints.  The
relation of a Chern-Simons theory on three-manifold with boundary to the
Wess-Zumino conformal field theory on the boundary has been exploited in doing
so.  Expectation value of Wilson link operators with
the same spin $n/2$
representation of $SU(2)$ living on each of the component knots of the link
yields a whole variety of link invaraints.  The Jones one variable polynomial
corresponds to $n=1$.  For higher $n$, these are the new link invaraints
discussed by Wadati, Deguchi and Akutsu from the point of view of $N=n+1$ state
exactly solvable statistical models$^{10}$.  As illustration of our method we
have
also computed explicitly these invariants for knots upto seven crossing points.
 The method can be generalized to links where different representations of
$SU(2)$ are placed on the component knots.  Such links may be called
multicoloured.  Expectation values of Wilson link operators associated with
such multicoloured links would then provide new link invariants.  The knowledge
of the expectation value of Wilson operators for any link with arbitrary
representations of the gauge group living on the component knots would then
provide a complete solution of the non-abelian Chern-Simons theory in three
dimensions.  A detailed discussion of this will be presented elsewhere.
\\
\par Generalization of our discussion to an arbitrary compact gauge group, say
$SU(N)$, is rather straight forward.  We shall take up
this  elsewhere.
\vskip2cm
\noindent{\bf Acknowledgements}
\\
We thank R.Jagannathan and K.Srinivasa Rao for discussions on quantum Racah
coefficients.

\vskip3cm
\noindent{\bf Appendix A}
\\
\par Here we shall list some useful properties of the quantum Racah
coefficients and
the duality matrix $a_{j\ell}$.
\\
\par
The quantum Racah coefficients are given by $^{11-13}$
\begin{eqnarray*}
\pmatrix {j_{1} & j_{2} & j_{12} \cr j_{3} & j_{4} & j_{23}}&= &
\Delta(j_1,j_2,j_{12}) \Delta(j_3,j_4,j_{12}) \Delta(j_1,j_4,j_{23})
\Delta(j_3,j_2,j_{23})  \\
&&\sum_{m\geq0}{(-)^{m} [m+1]!}
\Bigl\{ {[m-j_1-j_2-j_{12}]}!\Bigr. \\
&&{[m-j_3-j_4-j_{12}]}!
{[m-j_1-j_4-j_{23}]}!\\
&&{[m-j_3-j_2-j_{23}]}!
{[j_1+j_2+j_3+j_4-m]}! \\
&& \Bigl.{[j_1+j_3+j_{12}+j_{23}-m]}!{[j_2+j_4+j_{12}+j_{23}-m]}!\Bigr\}^{-1}
\hskip1cm (A.1)
\end{eqnarray*}
and
$$ \Delta(a,b,c) = \sqrt{{{[-a+b+c]![a-b+c]![a+b-c]!}\over{[a+b+c+1]!}}}
\eqno(A.2) $$
Here $[a]!= [a][a-1][a-2]...[2][1]$. The $SU(2)$ spins are related as
$\overrightarrow{j}_1 + \overrightarrow{j}_2 + \overrightarrow{j}_3 =
\overrightarrow{j}_4 , \overrightarrow{j}_1 + \overrightarrow{j}_2 =
\overrightarrow{j}_{12} , \overrightarrow{j}_2 + \overrightarrow{j}_3 =
\overrightarrow{j}_{23}$
\\
\par The duality matrix $a_{j\ell}$ where three spins, each $n/2$ is combined
into
spin $n/2$ is given by
$$ a_{j\ell}  =  (-)^{\ell+j-n} \sqrt{[2j+1][2\ell+1]} \left( \matrix {n/2 &
n/2 &
j \cr n/2 & n/2 & \ell } \right) \eqno(A.3) $$
The $q$-Racah coefficients satisfy the following properties$^{11}$ :
$$ \left( \matrix {j_1 & j_2 &
j \cr j_3 & j_4 & \ell} \right) =  \left( \matrix {j_1 & j_4 &
\ell \cr j_3 & j_2 & j } \right) \eqno(A.4) $$
$$\pmatrix {j_1 & j_2 &
0 \cr j_3 & j_4 & \ell }={(-1)^{\ell+j_{2}+j_{3}} \delta_{j_{1} j_{2}}
\delta_{j_{3} j_{4}} \over{\sqrt{[2j_2 +1][2j_3 +1]}}}\eqno(A.5) $$
$$ \sum_j [2j+1][2\ell +1] \pmatrix {j_1 & j_2 & j \cr j_3 & j_4 & \ell }
\pmatrix {j_1 & j_2 & j \cr j_3 & j_4 & \ell
'} = \delta_{\ell \ell '} \eqno(A.6) $$
\begin{eqnarray*}
 \sum_x (-)^{j+\ell+x} [2x+1] q^{-C_x} \pmatrix {j_1 & j_2 & x \cr j_3 & j_4 &
j}
\pmatrix {j_1 & j_2 & x \cr j_4 & j_3 & \ell } =
\ \ \ & &
\\  \pmatrix {j_1 & j_3 & j \cr j_2 & j_4 &
\ell } q^{C_j/2 + C_\ell /2}
q^{-C_{j_{1}}/2-C_{j_{2}}/2-C_{j_{3}}/2-C_{j_{4}}/2}
& & \hskip4cm  (A.7)
\end{eqnarray*}
where $ C_{j} = j(j+1)$ is the Casimir invariant
in the spin $j$ representation $R_{2j}$.
These relations respectively imply the following properties for the duality
matrix $a_{j\ell}$ defined in (A.3)
$$ a_{j\ell}  =  a_{\ell j}  \eqno(A.8)$$
$$ a_{jo}  =  {\sqrt{[2j+1]} \over{[n+1]}}  \eqno(A.9) $$
$$ \sum^n_{\ell=0} a_{\ell i} a_{\ell j}  =  \delta_{ij}  \eqno(A.10) $$
$$ \sum_x (-)^{n-x}  q^{\pm {1\over2}(n(n+2)-x(x+1))}
\  a_{jx} a_{\ell x} = (-)^{j+\ell}
q^{\pm {(j(j+1)+\ell(\ell+1))}/2} \ a_{j\ell}  \eqno(A.11) $$
\noindent
Further notice that (A.9) and (A.10) imply:
$$ \sum^n_{\ell=0} \sqrt{[2\ell+1]} \ a_{\ell j}  =  \delta_{jo} [n+1]
\eqno(A.12)
$$
and (A.9) and (A.11) imply :
$$\sum_{j=0}^n\sqrt{[2j+1]} \ (-)^j \
q^{\pm{j(j+1)\over2}} \
a_{jl}=(-)^{n-l}\sqrt{[2\ell +1]} \
q^{\pm{1\over2}(n(n+2)-\ell (\ell +1))}
\eqno(A.13) $$
Also (A.10) and (A.11) imply:
$$ \sum_{r,s} (-)^{r+s} q^{\pm({r(r+1)\over2}+{s(s+1)\over2})} \  a_{\ell r}
a_{rs}
a_{sj}  =  \delta_{\ell j} (-)^{n-\ell} \
q^{\pm({n(n+1)\over2}+{\ell(\ell+1)\over2})}
\eqno(A.14)$$
Another useful relation is
$$ \sum_{r,s} \sqrt{[2r+1][2s+1]} \  a_{rs} \  q^{r(r+1)+s(s+1)} = \sum
[2\ell+1] \
(-)^\ell  \ q^{{3\over2} \ell(\ell+1)}  \eqno(A.15)$$
\\
\par It is instructive to write down this duality matrix $a_{j\ell}$ explicitly
for
various low values of $n$.  We need this to compare the results obtained here
with those of Jones$^{2}$ and Akutsu, Deguchi and Wadati$^{10}$ for the link
invariants for $n=1$ and $n=2$ respectively.  The duality matrix for these two
values of $n$ reads :
\vskip1cm
\noindent{\bf (i) n = 1}
$$ a_{j\ell} = {1\over{[2]}} \pmatrix{ 1 & \sqrt{[3]} \cr
\sqrt{[3]} & -1}   \eqno(A.16) $$
\newpage
{\bf (ii) n = 2}

$$ a_{j\ell} = {1\over{[3]}} \pmatrix{1 & \sqrt{[3]} & \sqrt{[5]} \cr
\sqrt{[3]} & {{[3]([5]-1)}\over{[4][2]}} & - {{[2]\sqrt{[5][3]}}\over{[4]}} \cr
\sqrt{[5]} & -{{[2]\sqrt{[5][3]}}\over{[4]}} & [2]\over{[4]} }  \eqno(A.17) $$
\vskip2cm
\noindent{\bf Appendix B}
\\
\par Here we shall illustrate by an example how the framework developed in this
paper can be used to rederive many of the results already known for Jones
polynomials.
\\
\par
The generalization of numerator-denominator theorem of Conway for Jones
polynomials has been proved by Lickorish and Millet$^{16}$.  We shall present a
new proof for this
theorem here  .
\vskip1cm
\noindent{\bf Theorem :} For the links depicted in Fig.21,this theorem
states that
\begin{eqnarray*}
\cup(\cup ^2 -1){\hat L}_o(\hat A,\hat B)& =&
\cup\Big\{\hat L_o(\hat A,\hat Q_o^H)\hat L_o(\hat
Q_o^H,\hat B)+\hat L_o(\hat A,\hat Q^V_o)\hat
L_o(\hat Q^V_o,\hat B)\Big\} \\
& &-\Big\{\hat L_o(\hat A,\hat Q^H_o)
\hat L_o(\hat Q^V_o,\hat B)+\hat L_o(\hat A,\hat
Q^V_o)\hat L_o(\hat Q^H_o,\hat B)\Big\}\\
&& \hskip4in (B.1)
\end{eqnarray*}
where the symbol for each of the link diagrams itself represents the Jones link
invariant $V_{1}[L]$.
\\
\par
The proof of this theorem is rather straight forward in our framework.  We
write the link invariants above as :
$$ \hat{L_{o}}(\hat{A},\hat{B}) = \hat{\mu_{o}}(\hat{A}) \hat{\mu_{o}}
(\hat{B}) + \hat{\mu_{1}}(\hat{A}) \hat{\mu_{1}}(\hat{B})  \eqno(B.2) $$
$$ \hat{L_{o}}(\hat Q_o^V,\hat{B})  =  \hat{\mu_o}(\hat{B}) + \sqrt{[3]} \
\hat{\mu_1}(\hat{B}) \eqno(B.3) $$
$$ \hat{L_{o}}(\hat Q_o^H,\hat{B}) =  \hat{\mu_o}(\hat{B}) \  [2]  \eqno(B.4)
$$
$$ \hat{L_{o}}(\hat{A},\hat Q_o^V)  =  \hat{\mu_o}(\hat{A}) + \sqrt{[3]}
 \ \hat{\mu_1}(\hat{A}) \eqno(B.5) $$
$$ \hat{L_{o}}(\hat{A},\hat Q_o^H)  =  \hat{\mu_o}(\hat{A}) \  [2]  \eqno(B.6)
$$
where we have used from Theorem 9.
$$ \hat{\mu_\ell} (\hat Q_o^V)  =  \sqrt{[2\ell+1]} ,\
\  \ \ \hat{\mu_\ell}(\hat Q_o^H)
= \sum^n_{j=0} \sqrt{[2j+1]} \ a_{j\ell} = \delta_{\ell o}[n+1] $$
Now if we solve for $\hat{\mu_\ell}(\hat{A})$
and ${\hat\mu}_\ell(\hat{B}), \ \  \ell=0,1$ from Eqns(B.3-B.6), we have
$$ \hat{\mu_o}(\hat{X})  =  \cup \  {\hat
L}_0({\hat Q}_0^H,{\hat B})$$
$$ {\hat\mu}_1(\hat{X}) = {{\cup  \ {\hat
L}_0(\hat X,{\hat Q}_0^V) -{\hat L}_0({\hat X},{\hat
Q}_0^H)}\over {\cup  \ \sqrt{(\cup ^2-1)}}} \eqno(B.7) $$
where $\hat{X} = \hat{A},\hat{B}$ and the link invariant for the unknot
$(n=1)$ $\cup =[2]$ and $[3] = \cup^2-1$.  Substituting these values $(B.7)$
into
$(B.2)$ then, immediately yields the theorem above.
\vskip2cm
\noindent{\bf Appendix C}
\\
\par In the following, we list various normalized functional integrals expanded
in terms of the
bases vectors referring to the middle two  of the four strands attached
to each of the boundaries ,$S^{2}$.  The diagram for the manifold itself will
represent the normalised functional integral.In the following numbers on the
left
hand side refer to the diagrams in Fig.22,respectively:
\begin{eqnarray*}
(1) & = & \sum^n_{\ell = 0} \sqrt{[2\ell + 1]} \mid \hat \phi_\ell^{(1)}
\rangle \\
(2) & = & [n+1] \mid \hat \phi_o^{(1)} \rangle \\
(3) & = & \sum^n_{\ell=0} \sqrt{[2\ell+1]} {q^{m\ell(\ell+1)}} \mid
{\hat \phi_\ell^{(1)}} \rangle ,\\
(4) & = & \sum^n_{\ell=0} \sqrt{[2\ell+1]} (-) q^{{2m+1\over2} \ell(\ell+1)}
\mid
\hat \phi_\ell^{(1)} \rangle \\
(5) & = & \sum^n_{j,\ell=0} \sqrt{[2j+1]} a_{j\ell} q^{mj(j+1)} \mid
\hat \phi_\ell^{(1)} \rangle ,\\
(6) & = & \sum^n_{j\ell} \sqrt{[2j+1]} a_{j\ell} (-)^j q^{{2m+1\over2} j(j+1)}
\mid
\hat \phi_\ell^{(1)} \rangle \\
(7) & = & \sum_{j} q^{mj(j+1)} \mid \hat \phi_j^{(1)} \rangle \mid
\hat \phi_j^{(2)} \rangle \\
(8) & = & \sum^n_{j=0} (-)^j q^{{2m+1\over2} j(j+1)} \mid\hat \phi_j^{(1)}
\rangle \mid\hat \phi_j^{(2)} \rangle \\
(9) & = & \sum_{j,\ell} \sqrt {{[2\ell+1]} \over {[2j+1]}} a_{\ell j}
q^{m(n(n+2) - \ell(\ell+1))} \mid\hat \phi_j^{(1)} \rangle \mid\hat
\phi_j^{(2)} \rangle \\
(10) & = & \sum_{j,\ell} \sqrt {{[2\ell+1]} \over {[2j+1]}} a_{\ell j}
q^{m\ell(\ell+1)} \mid\hat \phi_j^{(1)} \rangle
\mid\hat \phi_j^{(2)} \rangle \\
(11) & = & \sum_{j,\ell = 0} \sqrt {{[2\ell+1]} \over {[2j+1]}} a_{\ell j}
(-)^{n-\ell}
q^{{2m+1\over2}(n(n+2)-\ell(\ell+1))} \mid\hat \phi_j^{(1)} \rangle
\mid\hat \phi_j^{(2)} \rangle \\
(12) & = & \sum \sqrt {{[2\ell+1]} \over {[2j+1]}} a_{\ell j} (-)^\ell
q^{{2m+1\over2} \ell(\ell+1))} \mid\hat \phi_j^{(1)} \rangle
\mid\hat \phi_j^{(2)} \rangle \\
(13) & = & \sum_{j\ell r} a_{jr} a_{\ell r} q^{(m+p)r(r+1)} \mid\hat
\phi_j^{(1)} \rangle
\mid\hat \phi_\ell^{(2)} \rangle \\
(14) & = & \sum_{j\ell r} a_{jr} a_{\ell r} q^{(m+p+1)r(r+1)} \mid\hat
\phi_j^{(1)} \rangle
\mid\hat \phi_\ell^{(2)} \rangle \\
(15) & = & \sum a_{jr} a_{\ell r} (-)^r q^{{({{2m+2p+1} \over2})}r(r+1)}
\mid\hat
\phi_j^{(1)} \rangle
\mid\hat \phi_\ell^{(2)} \rangle \\
(16) & = & \sum_{ij\ell r}  {a_{ij} a_{jr} a_{\ell r} \sqrt{[2\ell+1]} \over
\sqrt {[2r+1]}} q^{m\ell(\ell+1)} \mid \hat{\phi}_i^{(1)} \rangle \mid
\hat{\phi}_j^{(2)} \rangle  \\
(17) & = & \sum_{ij\ell r}  {a_{ir} a_{jr} a_{\ell r} \sqrt{[2\ell+1]} \over
\sqrt{[2r+1]}} q^{m(n(n+2)-\ell(\ell+1))} \mid \hat{\phi}_i^{(1)} \rangle \mid
\hat{\phi}_j^{(2)} \rangle \\
(18) & = & \sum_{ij\ell r}  {a_{ir} a_{jr} a_{\ell r} \sqrt{[2\ell+1]} \over
\sqrt {[2r+1]}} (-)^\ell q^{({{2m+1} \over2}) \ell(\ell+1)} \mid
\hat{\phi}_i^{(1)} \rangle \mid
\hat{\phi}_j^{(2)} \rangle \\
(19) & = & \sum_{ij\ell r}  {a_{ir} a_{jr} a_{\ell r} \sqrt {[2\ell+1]} \over
\sqrt {[2r+1]}} (-)^{n-\ell} q^{({{2m+1} \over2})(n(n+2)-\ell(\ell+1))} \mid
\hat{\phi}_i^{(1)} \rangle \mid
\hat{\phi}_j^{(2)} \rangle \\
(20) & = & \sum_{ij\ell r s}  {a_{ir} a_{jr} a_{\ell r} a_{sr} \sqrt{[2\ell+1]
[2s+1]} \over
{[2r+1]}} q^{m\ell(\ell+1)} q^{ps(s+1)} \mid \hat{\phi}_i^{(1)} \rangle \mid
\hat{\phi}_j^{(2)} \rangle \\
(21) & = & \sum_{ij\ell r s}  {a_{ir} a_{jr} a_{\ell r} a_{sr} \over
{[2r+1]}} \sqrt{[2\ell+1] [2s+1]} (-)^{\ell+s} q^{({{2m+1}
\over2})(n(n+2)-\ell(\ell+1))} \\  & & \hskip5cm  \bullet  \ q^{({{2p+1}
\over2})(n(n+2)-s(s+1))} \mid \hat{\phi}_i^{(1)} \rangle \mid
\hat{\phi}_j^{(2)} \rangle \\
(22) & = & \sum_{ij\ell r s}  {a_{ir} a_{jr} a_{\ell r} a_{sr}  \over
{[2r+1]}} \sqrt{[2\ell+1] [2s+1]}q^{m(n(n+2)-\ell(\ell+1))}\\  & & \hskip5cm
\bullet
 \ q^{p(n(n+2)-s(s+1))} \mid \hat{\phi}_i^{(1)} \rangle \mid
\hat{\phi}_j^{(2)} \rangle \\
(23) & = & \sum_{ij\ell r s}  {a_{ir} a_{jr} a_{\ell r} a_{sr} \sqrt{[2\ell+1]
[2s+1]} \over {[2r+1]}} (-)^{\ell+s} q^{({2m+1 \over2}) \ell(\ell+1)}
\\  & &  \hskip5cm \bullet  \ q^{({2p+1 \over2}) s(s+1)} \mid
\hat{\phi}_i^{(1)} \rangle \mid
\hat{\phi}_j^{(2)} \rangle \\
(24) & = & \sum_{j\ell r} a_{rj} a_{r\ell} q^{m(n(n+2)-r(r+1))} \mid
\hat{\phi}_j^{(1)} \rangle \mid \hat{\phi}_{\ell}^{(2)} \rangle \hskip5.5cm
(C.1)
\end{eqnarray*}
\\
\par  Now let us present some functional integrals expanded in terms of the
bases
vectors referring to the first two strands on the left (and equivalently the
last two
on the right) attached to the boundaries, each of which is an $S^2$ with four
punctures as shown in Fig.23.For the diagrams listed in this figure,the
normalized functional integrals are as follows :
\\
\begin{eqnarray*}
(1) & = & \sum^n_{\ell = 0} \sqrt{[2\ell+1]} \mid \phi_\ell^{(1)} \rangle, \\
(2) & = & \sum^n_{\ell = 0} \sqrt{[2\ell+1]} (-)^{m(n-\ell)}
q^{{m\over2}(n(n+2)-\ell(\ell+1))} \mid {\phi}_{\ell}^{(1)} \rangle \\
(3) & = & \sum^n_{j, \ell} \sqrt {[2j+1]} (-)^j a_{j \ell} q^{({2m+1 \over2})
j(j+1)} \mid {\phi_\ell^{(1)}} \rangle, \\
(4) & = & \sum^n_{j \ell} \sqrt{[2j+1]} a_{j \ell} q^{mj(j+1)} \mid
{\phi}_{\ell}^{(1)} \rangle \\
(5) & = & \sum^n_{\ell = 0} (-)^{(m+p)(n- \ell)}
q^{({m+p \over2}) (n(n+2)- \ell(\ell+1))} \mid {\phi_{\ell}^{(1)}} \rangle\mid
{\phi_{\ell}^{(2)}} \rangle \\
(6) & = & \sum_{j \ell} \sqrt{[2j+1]\over{[2\ell+1]}} (-)^j a_{j\ell}
q^{({{2m+1} \over2})j(j+1)}
\mid {\phi_{\ell}^{(1)}} \rangle\mid {\phi_{\ell}^{(2)}} \rangle \\
(7) & = & \sum_{j \ell} \sqrt{[2j+1]\over{[2 \ell+1]}} q^{mj(j+1)} a_{j \ell}
\mid {\phi_{\ell}^{(1)}} \rangle\mid {\phi_{\ell}^{(2)}} \rangle \\
(8) & = & \sum_{j \ell} \sqrt{[2j+1]\over{[2 \ell+1]}} q^{m(n(n+2)-j(j+1))}
a_{j
\ell} \mid {\phi_{\ell}^{(1)}} \rangle\mid {\phi_{\ell}^{(2)}} \rangle \\
(9) & = & \sum_{j \ell} \sqrt{[2j+1]\over{[2 \ell+1]}} (-)^{n-j}  q^{({{2m+1}
\over2})(n(n+2)-j(j+1))} a_{j\ell} \mid {\phi_{\ell}^{(1)}} \rangle\mid
{\phi_{\ell}^{(2)}} \rangle \\
(10) & = & \sum_{ij \ell} q^{m \ell(\ell+1)} a_{\ell i} a_{\ell j} \mid
{\phi_{i}^{(1)}} \rangle\mid {\phi_{j}^{(2)}} \rangle \\
(11) & = & \sum_{ij \ell} (-)^\ell q^{({{2m+1} \over2}) \ell(\ell+1)} a_{\ell
i}
a_{\ell j} \mid {\phi_{i}^{(1)}} \rangle\mid {\phi_{j}^{(2)}} \rangle \\
(12) & = & \sum_{ij \ell r}  {a_{ir} a_{jr} a_{\ell r} \sqrt{[2 \ell+1]}
\over{\sqrt{[2r+1]}}}
q^{m \ell(\ell+1)}  \mid {\phi_{i}^{(1)}} \rangle\mid {\phi}_{j}^{(2)} \rangle
\\
(13) & = & \sum_{ij \ell r}  {{a_{ir} a_{jr} a_{\ell r} \sqrt{[2 \ell+1]}}
\over{\sqrt{[2r+1]}}} (-)^{\ell} q^{({{2m+1} \over2}) \ell(\ell+1)}
\mid {\phi_{i}^{(1)}} \rangle\mid {\phi_{j}^{(2)}} \rangle \\
(14) & = & \sum_{ij \ell r}  {a_{ir} a_{jr} a_{\ell r} \sqrt{[2 \ell+1]}
\over{\sqrt{[2r+1]}}} q^{m(n(n+2)-\ell(\ell+1))}
\mid {\phi_{i}^{(1)}} \rangle\mid {\phi_{j}^{(2)}} \rangle \\
(15) & = &\sum_{ij \ell r}  {a_{ir} a_{jr} a_{\ell r} \sqrt{[2 \ell+1]}
\over{\sqrt{[2r+1]}}} (-)^{n- \ell} q^{({{2m+1} \over2})(n(n+2)- \ell(\ell+1))}
\mid {\phi_{i}^{(1)}} \rangle\mid {\phi_{j}^{(2)}} \rangle \hskip2cm  (C.2)
\end{eqnarray*}

\newpage
\noindent{\bf Figure Captions}
\vskip1cm
Fig.[1]. Connected sum of two links $L_1,L_2$.
\vskip1cm
Fig.[2]. Examples of rooms.
\vskip1cm
Fig.[3]. Composition of two balls $B_1,B_2$ leads to
link $L_m(A)$ in $S^3$.
\vskip1cm
Fig.[4]. Half twists in two strands.
\vskip1cm
Fig.[5]. Composing of balls $B_1$ and $B_2$ gives
link ${\hat L}_{2m}(\hat A)$ in $S^3$.
\vskip1cm
Fig.[6]. Link ${\hat L}_{2m+1}({\hat A}')$ with room
\begin{picture}(10,7)
\put(5,1.5){\circle{7}}
\put(3,0){$\hat A'$}
\put(3,4){\vector(0,1){3}}
\put(7,7){\vector(0,-1){3}}
\put(7,-4){\vector(0,1){3}}
\put(3,-1){\vector(0,-1){3}}
\put(15,0){.}
\end{picture}
\vskip1cm
Fig.[7]. Composition of balls $B_1$ and $B_2$ yields
link $L(A_1,A_2)$ in $S^3$.
\vskip1cm
Fig.[8]. (a)Parallely oriented two-strand braid with
m half-twists and \\
 \hskip2cm (b) its closure ${\cal L}_{m}$.
\vskip1cm
Fig.[9]. Composition of balls $B_1$ and $B_2$ yields
link ${\hat L}_{2m}(\hat A_1,\hat A_2) $ in
$S^3$.
\vskip1cm
Fig.[10].(a)Oppositely oriented two-strand braid
with 2m half-twists and \\
\hskip2cm  (b)its closure $\hat
{\cal L}_{2m}$.
\vskip1cm
Fig.[11]. Rooms $Q^V_m$ and $Q^H_{2p+1}$.
\vskip1cm
Fig.[12]. Rooms $\hat Q^V_m$,$\hat Q^H_{2p}$ and
$\hat
{Q'}^H_p$
\vskip1cm
Fig.[13]. The links (a)$\hat L_o(\hat Q^H_{2m},\hat
Q^H_{2p})\equiv \hat {\cal L}_{2m+2p}$   and \\
\hskip2cm (b)$\hat L_o(\hat Q^H_{2m},\hat Q^V_{2p})\equiv
\hat L_o(\hat Q^V_{2m},\hat Q^H_{2p})$
\vskip1cm
Fig.[14]. The links (a)$L_o(Q^H_{2p+1},Q^H_{2m+1})=\hat
{\cal L}_{2p+2m+2}$ and  \\  \hskip2cm (b)$L_o(Q^H_1,Q^V_m)=
{\cal L}_{m+1}$
\vskip1cm
Fig.[15]. Vectors corresponding to Eqns.7.1 and 7.2
respectively.
\vskip1cm
Fig.[16]. Diagrammatic representations of functional integrals
$\hat {\nu}_1,\hat {\nu}_2,\hat {\nu}_3 $ \\
\hskip2cm in
Eqns.7.3-5  respectively.
\vskip1cm
Fig.[17]. Functional integral over a manifold with r
boundaries ,$\hat {\nu}_r$ of Eqn.7.6.
\vskip1cm
Fig.[18]. Functional integral over a manifold with r
boundaries,${\nu}_r$ of Eqn.7.9.
\vskip1cm
Fig.[19]. Functional integral over a manifold with
$2r+2$ boundaries (Eqn.7.12).
\vskip1cm
Fig.[20]. Knot projections upto seven crossing
points.
\vskip1cm
Fig.[21]. Link diagrams for numerator-denominator theorem
(Eqn.B.1).
\vskip1cm
Fig.[22]. Building blocks corresponding to the
Eqns.(C.1) respectively.

\vskip1cm
Fig.[23].Building blocks corresponding to the
Eqns.(C.2) respectively.


\begin{thebibliography}{99}
\bibitem{19.} Our normalization differs from those in Refs.2 and 10, where the
	invaraint for an unknot is taken to be one.  Thus to compare our results
	for $n=1$ and $n=2$ with those of these references, we need to multiply
	their expressions by $V_{n}[\cup] = [n+1]$.
\end{thebibliography}
\end{document}